# Convexity in scientific collaboration networks

Lovro Šubelj[a, *]

Dalibor Fiala[b]

Tadej Ciglarič[a]

Luka Kronegger[c]

[a] University of Ljubljana, Faculty of Computer and Information Science

Večna pot 113, 1000 Ljubljana, Slovenia

[b] University of West Bohemia, Department of Computer Science and Engineering

Univerzitní 8, 30614 Plzeň, Czech Republic

[c] University of Ljubljana, Faculty of Social Sciences

Kardeljeva ploščad 5, 1000 Ljubljana, Slovenia

* Corresponding author. Tel.: +386 1 479 8233.
Email addresses: lovro.subelj@fri.uni-lj.si (L. Šubelj), dalfia@kiv.zcu.cz (D. Fiala), tc2922@student.uni-lj.si (T. Ciglarič), luka.kronegger@fdv.uni-lj.si (L. Kronegger).

**Abstract**: Convexity in a network (graph) has been recently defined as a property of each of its subgraphs to include all shortest paths between the nodes of that subgraph. It can be measured on the scale [0, 1] with 1 being assigned to fully convex networks. The largest convex component of a graph that emerges after the removal of the least number of edges is called a convex skeleton. It is basically a tree of cliques, which has been shown to have many interesting features. In this article the notions of convexity and convex skeletons in the context of scientific collaboration networks are discussed. More specifically, we analyze the co-authorship networks of Slovenian researchers in computer science, physics, sociology, mathematics, and economics and extract convex skeletons from them. We then compare these convex skeletons with the residual graphs (remainders) in terms of collaboration frequency distributions by various parameters such as the publication year and type, co-authors' birth year, status, gender, discipline, etc. We also show the top-ranked scientists by four basic centrality measures as calculated on the original networks and their skeletons and conclude that convex skeletons may help detect influential scholars that are hardly identifiable in the original collaboration network. As their inherent feature, convex skeletons retain the properties of collaboration networks. These include high-level structural properties but also the fact that the same authors are highlighted by centrality measures. Moreover, the most important ties and thus the most important collaborations are retained in the skeletons.

## 1.    Introduction

The notion of network convexity has been defined only recently by Marc and Šubelj (2018). By their definition, a *convex network* is a connected (undirected) graph where every connected subgraph includes all shortest paths between its nodes. Such a network takes generally the form of a tree of cliques. Furthermore, by removing the least number of edges from a graph to obtain its largest convex subgraph, we extract a *convex skeleton* of the original graph with many interesting properties (Šubelj, 2018). One of these properties is the fact that the resulting convex skeleton is a generalized spanning tree with potential cliques retained, which leads to a network structure with a similar degree distribution, clustering, and node distances like in the original graph but with unique shortest paths between nodes. Convex skeletons can thus be regarded as a network abstraction technique with possible applications in network modelling and visualization. Particularly for scientific collaboration networks since they turn out to be rather convex (Marc and Šubelj, 2018). The goal of the present study is to build upon the two above analyses and investigate the concept of convexity and convex skeletons in the context of scientific collaboration (or co-authorship) networks at the level of authors (i.e. here we are not interested in country-level or institutional collaboration at all). The questions we would like to answer by this analysis is whether convex skeletons extracted from author collaboration networks can be used as their meaningful abstraction, whether author rankings by centrality measures calculated on convex skeletons differ substantially from those calculated on the original networks, and what kind of edges are removed when convex skeletons are extracted.

The motivation for studying convexity in co-authorship networks is at least threefold. Firstly, as already mentioned above, different collaboration networks turn out to be rather convex (Marc and Šubelj, 2018; Šubelj, 2018), which is in contrast to paper citation and other bibliographic networks. Convex skeletons should therefore represent their meaningful abstraction, which is not the case for the latter. Secondly, since co-authorship networks are projections of bipartite graphs of authors and papers, these are unions of cliques by construction. Yet, how these cliques are connected together is unknown. Convex skeletons propose possibly the simplest configuration in which cliques are connected in a tree and thus provide a very simple framework for understanding the structure of scientific co-authorship. Lastly, although convexity is a well understood property of different mathematical objects, and a key component in mathematical optimization, it has not been explored in networks until only recently. Different collaboration networks likely represent the most obvious setting to bridge this gap.

The article is organized in the following way. After recalling the published literature on collaboration networks in Section 2, we introduce the methods dealing with network convexity and convex skeletons in Section 3, and present the data underlying our analysis in Section 4: co-authorship networks of Slovene researchers in computer science, physics, sociology, mathematics, and economics. We extract convex skeletons from these collaboration networks and, due to space limitations, show the main results for the first three disciplines in Section 5 with a special focus on the differences between the skeleton and the residual network. (A residual network or remainder consists of the same set of nodes as the original network and of a subset of edges from the original network after the edges forming a convex skeleton have been removed). We then take conclusions, discuss the limitations of this study and propose future research directions in Section 6.

## 2. Related work

*2.1. Collaboration and co-authorship*

Due to the lack of collaboration-based network convexity studies, we will now only recall some fundamental and recent publications on scientific collaboration (or co-authorship) networks. One of the reasons to study research collaboration is the reported positive correlation between impact and collaboration (Hsu and Huang, 2011) and the review by Bozeman and Boardman (2014) is useful in investigating scholarly collaboration as such, but there are a number of other publications as well. Even though by far not all forms of research collaboration result in co-authorship (Laudel, 2002) and co-authorship is actually one of multiple components of research collaboration (Katz and Martin, 1997; Melin and Persson, 1996), for the purpose of this study we will consider those two terms as equivalent and interchangeable. In more detail, Laudel identified six types of research collaboration: collaboration involving a division of labour, service collaboration, provision of access to research equipment, transmission of know-how, mutual stimulation and trusted assessorship. Only the first type resulted systematically in co-authorships while the others were rewarded in acknowledgements (about a third of the collaborations inspected) or not at all (about a half). This last result contradicts the finding by Melin and Persson that (in their small-scale study of a single university) only around 5% of collaborations do not produce a co-authorship. Katz and Martin further admit that it is indeed difficult to define collaboration exactly and, unlike most others, even suggest that scientific collaboration brings about benefits as well as costs.

Hara et al. (2003) discuss types of collaboration and various factors that affect collaboration. These factors include compatibility, work connections, incentives, and socio-technical infrastructure. It is concluded that they sometimes facilitate collaboration and other times they hinder it. The effects of geographical distance on research collaboration have been studied by Katz (1994) and the geodesic distance of highly productive scientists in co-authorship networks has been investigated by Kretschmer (2004). The former study has analyzed intra-national collaborations between universities in the United Kingdom, Canada, and Australia and has found that the intensity of cooperative research decreases exponentially with the distance between the collaborating institutions. The latter has focused on the relationship between the average of geodesic distances (shortest paths) in a co-authorship network of a physics subfield and the productivity of researchers. It is concluded that more productive authors have, on average, a shorter geodesic distance to all other authors than less productive authors do. Regarding convexity, it is also defined using shortest paths, but unlike the above publications, it deals with the inclusion of shortest paths and not their length or number.

*2.2. Application fields*

Collaboration network analysis has been carried out for individual scientific disciplines (Kim and Diesner, 2016; Ding, 2011; Franceschet, 2011; Liu et al., 2005), individual countries (Perc, 2010), both (Leifeld et al., 2017), or a bulk of countries (Glänzel, 2001). Kim and Diesner concentrated on the issue of disambiguating author names in six co-authorship networks with tens of thousands of nodes each. The scientific disciplines of interest were biology, computer science, nanoscience, neuroscience, physics, and a multidisciplinary field. They found that a disambiguation by given names' initials deflated some statistical properties of the networks such as the average of shortest paths or the clustering coefficient and inflated some others such as the mean number of collaborators per author or the largest component size. As we pointed out earlier, our research is concerned with shortest paths, but not with their quantitative properties. Ding analyzed a collaboration network of more than 20,000 information retrieval researchers and defined a new collaborative distance measure called the Salton number denoting the shortest path length between an author and this well-known scholar in the field.

A massive computer science co-authorship network with more than 680,000 nodes was explored by Franceschet. In addition to the standard paper co-authorship, conference and journal co-authorship networks with edges connecting authors publishing at the same venue were analyzed too. Another important aspect was the study of the temporal evolution of these networks' properties over several decades. Liu et al. examined a much smaller co-authorship network with some 1,500 authors in the field of digital libraries. Apart from investigating traditional network properties, they also applied enhanced centrality measures to a directed co-authorship network in order to determine the most influential authors. Perc investigated the development of the collaboration network of Slovenia's researchers during a 50-year period using the same data source as in this study (SICRIS – see later). The co-authorship network analyzed consisted of more than 7,000 authors in the later stages and its growth was shown to be guided by preferential attachment. The collaboration landscape of German political science was mapped by Leifeld et al. who inspected a co-authorship network of about 1,300 researchers stemming from joint activities within five years. Besides identifying the largest collaborative clusters of researchers and their topics, they also employed centrality measures to determine the most central scientists. Glänzel's paper is a pioneering study on the patterns of collaboration between individual countries. His analysis of scientific collaboration at the macro level included 50 most productive countries and can be considered an incentive for the investigation of research co-authorships at lower levels, including individual scholars.

*2.3. Evolution of co-authorship networks over time*

Additionally, Barabási et al. (2002) and, more recently, Kim and Diesner (2015; 2017) have explored the evolution of collaboration networks over time and indicators based on co-authorship networks for the evaluation of interdisciplinary research have been reviewed by Wagner et al. (2011). Barabási et al. studied two co-authorship networks in mathematics and neuroscience with hundreds of thousands of nodes and analyzed their dynamics and topological structure in the period 1991-1998. The networks were found to be scale-free (following a scale-free power-law degree distribution) and their evolution to be governed by preferential attachment. The average degree increased and the node separation decreased during the period. In their first paper (2015), Kim and Diesner examined a collaboration network in the field of information systems and made use of three different author name disambiguation techniques. Then they closely observed the evolution of several key network properties between 1984 and 2013 and concluded that the differences between the disambiguation methods amplified over time and that the selection of a proper disambiguation algorithm could largely influence the results of any collaboration network analysis. In this respect, our research presented in this paper is very robust because it does not include disambiguation (see Section 4).

In their second paper (2017), Kim and Diesner dealt with three domain-specific co-authorship networks (computer science, physics, and biomedicine) and one nation-wide collaboration network (Korea), each covering a period of almost twenty years. They defined a new time-aware measure of the likelihood of two authors writing a joint publication if they had a common co-author, which was shown to be considerably lower than the existing indicators of the same sort. Wagner noted that co-authorship networks were no more appropriate to determine the degree of research interdisciplinarity since department-level affiliations were often missing or outdated and finding out a researcher's speciality directly in papers or elsewhere was a tedious task. Ferligoj et al. (2015) modelled complete national networks of researchers using stochastic actor-oriented modelling, Huang et al. (2008) proposed a model for the evolution of a computer science collaboration network and, finally, an overview of dynamic co-authorship network analysis is given in Mali et al. (2012).

*2.4. Strength of links and centrality measures*

Our research involves both edge weighting and centrality measures like in the other studies below. The problem of measuring the strength of links in co-authorship networks has been addressed by Brandão and Moro (2017) and, last but not least, collaboration networks have also been exploited to determine the importance of scholars (Dehdarirad and Nasini, 2017; Fiala, 2013; Yan and Ding, 2009). Brandão and Moro introduced a new metric called "tieness", which was a combination of neighbourhood overlap and collaboration frequency. They applied it to measure the strength of collaborations and to determine strong and weak ties in computer science, medicine, and physics. A small co-authorship network of neuroscience articles was studied by Dehdarirad and Nasini who broke down the original bipartite author-paper network into two one-mode networks – paper-paper and author-author – and used the former to evaluate the nodes in the latter.

Fiala employed several degree and eigenvector centrality measures to assess the importance of researchers in a large collaboration graph of more than 1.2 million computer scientists. Prior to the computation, the undirected co-authorship network was transformed into a directed graph by replacing each undirected edge with two directed edges pointing in opposite directions. A similar approach was chosen by Yan and Ding, albeit for a much smaller network. They made use of four different centrality measures to a co-authorship network of about 10,000 library and information science researchers spanning a 20-year period. They discovered that the author rankings produced in this way were somewhat correlated with rankings by citations.

*2.5. Newman's analyses of scientific collaboration networks*

A special place and a starting point in the analysis of co-authorship networks for many studies (and the present paper is no exception) is the abundant work by M. E. J. Newman. Newman (2001a) has studied a variety of statistical properties of several scientific collaboration networks in biomedicine, physics, and computer science whose sizes ranged from about 10,000 to a million authors. The statistics investigated included the mean number of papers per author, the mean number of authors per paper, the mean number of collaborators per author, the size of the largest connected component, and the clustering coefficient. Later (Newman, 2001b), shortest paths and centrality measures were also inspected and the mean and maximum distances were calculated for each of the networks. Overall, it was found that the smallest mean number of authors per paper was in the computer science collaboration network that also had the smallest largest connected component but the biggest mean distance between authors and the largest clustering coefficient. By contrast, the largest mean number of authors per paper was discovered in high-energy physics with the smallest mean distance between authors.

As for biomedicine, it had the biggest largest connected component and the smallest clustering coefficient implying a lower probability for two co-authors on a third co-author's paper to also write a paper together. A summary of the above statistical properties of all four networks has been given in yet another article on the structure of collaboration networks (Newman, 2001c) along with distribution plots of the number of collaborators and papers in computer science and three subfields of physics. Again, the different structure of collaboration in the scientific disciplines under study was made apparent. Furthermore, an additional network of co-authorships in the field of mathematics with about 250,000 authors has been analyzed by the same researcher (Newman, 2004). Mathematics turned out to have an even lower mean number of authors per paper, a smaller mean distance between authors, and a smaller clustering coefficient than computer science, but a bigger largest connected component. The degree assortativity coefficient was found to be positive for all scientific fields under investigation suggesting that frequently collaborating authors collaborated with other frequently collaborating authors.

There have also been studies by other scholars concerned with cliques in research collaboration networks (e.g. Krumov et al., 2011; Kumar and Jan, 2014; Barbosa et al., 2017), but cliques are always convex as can be other subgraphs too, which will be discussed in the next section.

## 3. Methods

### 3.1. Convexity in networks

Convexity is a property of a part of a mathematical object that includes all the shortest paths between its units (Van de Vel, 1993). In the case of networks or graphs, the part is a subgraph and the units are the nodes of the network (Harary and Nieminen, 1981; Farber and Jamison, 1986). Hence, a subgraph induced on a subset of nodes $S$ is convex if it includes all the shortest paths or geodesics between the nodes in $S$ (i.e. paths through the least number of edges). For instance, any complete subgraph or a clique is convex and any connected subgraph of a tree, which is a subtree, is also convex. Note that any convex subgraph $S$ must obviously be a connected induced subgraph including all the edges between the nodes in $S$. See Figure 1 for examples of convex and non-convex graphs.

Insert Figure 1 here.

The extent to which a connected network is convex can be calculated using a global measure of network convexity $X$ (Marc and Šubelj, 2018), which is defined in the following way:

$$X = 1 - \sum_{t=1}^{n-1} \max\left(s(t) - s(t-1) - \frac{1}{n}, 0\right). \tag{1}$$

Here, $n$ is the number of nodes in the network and $s(t)$ is the average fraction of nodes in $S$ after $t$ expansions steps of a convex subgraph expansion algorithm (see below). $s(t) - s(t-1)$ in Eq. (1) is the increase in the fraction of nodes in $S$ at step $t$, which is compared to the increase in a fully convex network $1/n$. Therefore, convexity $X$ always falls within the interval [0, 1] with a value close to 1 indicating a highly convex network with a tree-like or clique-like structure such as a co-authorship network.

Eq. (1) relies on the outputs $s(t)$ of a greedy algorithm which is described in detail by Marc and Šubelj (2018). In brief, the procedure first chooses a random node in the graph for the set of nodes $S$, $t = 0$, and then expands $S$ by adding one more node in each further step $t \geq 1$ by following a random edge outside of $S$ (see Figure 2). In order for $S$ to induce a convex subgraph, it is expanded to its convex hull at the end of each step $t$. The convex hull of $S$ is the smallest convex subgraph including the nodes in $S$, which is uniquely defined (Harary and Nieminen, 1981). The algorithm terminates when all $n$ nodes are in $S$.

Insert Figure 2 here.

Obviously, $s(0) = 1/n$ and $s(1) = 2/n$, but $s(2) \geq 3/n$ and generally it holds that $s(t) \geq (t+1)/n$. More specifically, it has been shown by Marc and Šubelj (2018) that $s(t) \approx (t+1)/n$ in convex networks and $s(t) \gg (t+1)$ in non-convex networks, with $s(t)$ being the average of 100 runs of the expansion algorithm. In more simple terms, convex subgraphs found by the algorithm expand, and $s(t)$ thus increases, very slowly throughout the entire procedure in convex networks, but only in the first few steps in non-convex networks. After these first few steps, the expansion explodes and convex subgraphs include almost all nodes in the network with $s(t)$ close to 1. Consequently, convexity $X$ of such non-convex networks (e.g. random graphs) is near 0.

*3.2. Convex skeletons of networks*

A convex skeleton is defined as the largest part of a network such that every connected subset of nodes likely induces a convex subgraph (Šubelj, 2018). Therefore, it is the largest high-convexity backbone of a network. A fully convex skeleton would be a collection of cliques stitched together in a tree arrangement and thus a tree of cliques, where any two cliques can overlap in at most one node. It can be regarded as a generalization of a network spanning tree in which each edge can be replaced with a clique of arbitrary size. As a consequence, every connected network has at least one convex skeleton, which is a spanning tree, but most probably many more.

A polynomial-time algorithm for the discovery of the largest high-convexity parts of networks (i.e. convex skeletons) by a targeted removal of the least number of edges has been described by Šubelj (2018). In this study, we use the iterative algorithm provided by Ciglarič (2017) that removes a single edge at a time by maximizing the network clustering coefficient (Newman et al., 2001), while ensuring that the resulting skeleton remains connected. As an example, Figures 3 and 4 show the wiring diagram of a particular realization (1 out of 100) of a convex skeleton of the Slovenian computer science co-authorship network (with researchers' names and their unique IDs in the SICRIS database in brackets).

Insert Figure 3 here.

Insert Figure 4 here.

It has been shown by Šubelj (2018) that unlike spanning trees (see Figure 5) convex skeletons extracted from networks retain their basic topological properties like the degree distribution, clustering, connectivity, and also distances between nodes. The shortest paths between the nodes in a convex skeleton are also largely unique. Moreover, the skeleton of a small co-authorship network of the Slovenian computer scientists retained the strongest ties between authors. In this respect, convex skeletons may be regarded as a network simplification or backboning technique with possible applications in network abstraction, visualization, sampling, modelling, comparison, and others.

Insert Figure 5 here.

## 4. Data

For the experiments reported in this study, we used data from the SICRIS (www.sicris.si) database. SICRIS stands for Slovenian Current Research Information System and is a database of the Slovenian Research Agency serving primarily the purpose of research evaluation (Rodela, 2016). Every registered Slovenian researcher (out of about 15,000) is assigned a unique identifier and categorized into one or two scientific fields (*Natural sciences and mathematics*, *Engineering sciences and technologies*, *Medical sciences*, *Biotechnical sciences*, *Social sciences*, *Humanities*, and *Interdisciplinary studies*). Researchers' publication records are input on demand in the database by its administrators and when processing the data extracted from SICRIS, there is no need to disambiguate author names or to identify authors' research fields. The co-authorship networks investigated in this analysis were derived from SICRIS publication records, considered as scientific publications by the national research agency, spanning the period 1960-2010.

The three main scientific collaboration networks analyzed of computer science, physics, and sociology researchers had 475, 425 and 145 nodes, respectively, and 1,548, 2,223, and 596 edges, respectively, as summarized in Tables 1, 2, and 3. In addition, for the verification of our findings we also inspected another two co-authorship networks of mathematicians and economists with 167 and 414 nodes and 349 and 1,386 edges, which are shown in Tables A.1 and A.2 too for the sake of completeness. However, we will not discuss these two networks further in detail because the results based on them did not differ significantly from the main outcomes achieved from the first three co-authorship networks. Tables 1, 2, and 3 show some structural properties of each network as well as of the corresponding convex skeleton extracted from it, such as the number of nodes and edges, fraction of nodes in the largest connected component, mean degree of nodes, mean distance between nodes, assortativity, clustering coefficient, and convexity. Regarding the respective convex skeletons, their convexity is of course higher than that of the original network, together with the average distance and assortativity, while the average degree and clustering both decrease in them[1]. The total edge (co-authorship) weight decreases to 51% (from 4,269), 66% (from 6,641), and 57% (from 836.8) in the skeletons of computer science, physics, and sociology, compared to the decrease to 44%, 42%, and 52% in the number of edges in the skeletons. That means that high-weight edges tend to be retained in the skeletons. This feature, however, is not present in economics. As for the other three backbone structures (maximum spanning tree and backbones consisting of the same number of high-betweenness or high-embeddedness edges as in the convex skeleton), unsurprisingly, the maximum spanning tree has the largest connected component and convexity but the worst assortativity and clustering.

Insert Table 1 here.

Insert Table 2 here.

Insert Table 3 here.

## 5. Results and discussion

We will now present the results of the identification of a convex skeleton in the co-authorship networks of computer science, physics, and sociology researchers. There will thus be three plots with 16 charts, each representing the number of co-authorships (weighted edges using fractional counting, see below) on the Y-axis according to various criteria on the X-axis in one of the three collaboration networks. The light (green) bars in the charts represent the data of the convex skeleton extracted from the graph and the dark (grey) ones depict the data of the remainder (or residual) graph, i.e. a graph with convex skeleton edges removed. In the construction of co-authorship networks, the links between authors can either be counted fully (i.e. with weight 1 each) or fractionally (with weight inversely proportional to the number of co-authors of the paper that produces collaborations). These two concepts are discussed by Perianes-Rodriguez et al. (2016). The fractional counting scheme itself has a subvariant (Leydesdorff and Park, 2017), which we call partial counting. We will first present the results achieved with fractional counting and then those obtained by using full and partial counting.

---

[1] In fact, there are two versions of the clustering coefficient (Newman, 2010). Different versions are used in the algorithm and in the results.

*5.1. Computer Science*

Figure 6 shows the distributions of various data in the computer science network. For instance, the distribution of publication years is depicted in the top left chart. As we can see, the skeleton distribution peaks shortly before 2005 with about 150 co-authorships whereas the remainder distribution's peak is shifted to more recent years and culminates before 2010 with roughly the same number of edges. It thus appears that during the convex skeleton identification process some newer collaborations were discarded. This is a phenomenon different from physics (see Figure 7), but similar to sociology (see Figure 8), about the reasons of which we can only speculate. They have likely to do with different co-authorship patterns in various scientific fields. What is almost certain, however, is the fact that those discarded collaborations had low strength and were connecting mostly remote strongly-tied communities (see Section 5.5). Regarding the other three top charts dealing with the number of co-authorships by paper types, prime papers, and points assigned to the papers published, there seems to be no striking difference between the skeleton and the remainder. Most papers are conference proceedings papers (type 1.08 followed by 1.01 for journal articles), are not top-tier journal articles (so-called "prime papers"), and are awarded no points in the research evaluation system[2].

Insert Figure 6 here.

As far as the scientific age of researchers is concerned (measured by their academic birth year – the year their first paper was published), the picture is similar to that of paper publication years. The convex skeleton favours older scholars with a peak around birth year 1980 and the remainder prefers younger ones with the most of them being born around 1990. Another interesting aspect is the age difference between two collaborating authors whose distribution is plotted in the second chart from the left in the second row of charts in Figure 3. Here the most co-authorships (well over 200) occur between researchers of the same age and a difference up to about 20 years (presumably the gap between a doctoral advisor and his student) is still quite common. The frequency of collaborations between authors whose age difference is bigger than 20 or 25 years declines sharply in both the skeleton and the remainder. There are no significant differences either in the (seniority) status and gender of the collaborating authors in the skeleton and the remainder. Most collaborators have the same (seniority) status and the same gender.

---

[2] We believe that the indicators presented in Figure 6 are self-explanatory. Additional information about them can be found in SICRIS (www.sicris.si).

As for the third row of charts in Figure 6, the only big difference between the skeleton and the remainder is the distribution of co-authorships by the number of papers written by the collaborating authors. In both the skeleton and the remainder, the most collaborations take place between scientists whose total publication count is about 80. But in the skeleton there is a second (lower) peak in the distribution at around 220. This second culmination is completely missing in the remainder, which is, however, consistent with a previous finding that the remainder prefers younger scholars who are, therefore, generally less productive and have fewer publications. The second chart in the third row deals with the difference in the number of papers written by two collaborating authors. The most collaborations occur between authors that have the same productivity or do not have a production difference greater than approximately 100 papers. (Again, this number may be a threshold difference between the production of a doctoral advisor and a doctoral student.) After the difference of about 100 papers, the frequency of collaboration begins to decline with a slight tendency of the skeleton to exhibit larger differences. The two remaining plots are almost the same for both the skeleton and the remainder – most collaborations take place within scientific fields and between authors who hold a doctoral degree.

Regarding the bottom charts in Figure 6, the only significant difference between the convex skeleton and the remainder is in the left-most chart showing the distribution of collaborations based on the number of points achieved by the co-authors in the Slovenian research assessment system. Similarly to the number of papers, the skeleton has two peaks: one shared with the remainder at a little less than 1000 points and another smaller one at 1800 points, which is practically absent in the remainder. This is in accordance with the observation that more senior (and thus more productive) researchers have their collaborations retained in the skeleton and removed in the remainder. The difference in points behaves in a similar way as that in papers, i.e. the most collaborations materialize between scholars with the same number of points and after the difference exceeds a certain threshold (around 1000), the collaboration frequency decreases in the skeleton as well as in the remainder. The last two charts document that the most co-authorships occur between scientists that have no prime papers at all and, therefore, the most frequent difference in the number of prime papers is zero as well.

A quick look at Table S.1 in the electronic supplement with top 20 computer scientists sorted by four centrality measures (degree, PageRank[3], betweenness, and closeness) reveals that most prominent researchers in the original co-authorship network are ranked high by several indicators and only seven of them are appreciated by one: Milan Ojsteršek [6823] and Matjaž B. Jurič [18337] by degree, Aleš Leonardis [5896] by PageRank, Andrej Dobnikar [2272] and Saša Divjak [2268] by betweenness, and Bojan Cestnik [5806] and Viljan Mahnič [3307] by closeness. This is quite different from the top rankings based on the convex skeleton where the number of unique researchers is much higher (19). And while there are authors who are at the top both in the network and the skeleton by any metric like Matjaž Gams [8501], Marjan Krisper [1697], Marjan Heričko [11064], Ivan Rozman [8067], Ivan Bratko [2275], or Vladislav Rajkovič [1074], there are others like Marjan Mernik [11191] who appears in the top 20 in the network by any measure but nowhere in the skeleton or Bruno Stiglic [3034] who is highly ranked in the skeleton (except betweenness) but not in the network. We may thus assume that the latter researcher, by his influential position in the tree of cliques (convex skeleton) of research collaborations, probably reaches out to more isolated scholars. For reference, the whole computer science network is visualized in Figure 3.

---

[3] PageRank on an undirected graph is computed by a power iteration algorithm by replacing each undirected edge with two directed edges pointing in opposing directions.

*5.2. Physics*

The same way as above, distributions in the co-authorship network of physicists are depicted in Figure 7. Unlike computer science, the most papers in the physics skeleton appear only after 2005, which is roughly 10 years later than the publication peak of the remainder. Also the prevailing paper type here are journal articles and there are around twice as many of them in the skeleton than in the remainder. This is corroborated by a similar number of primary and non-primary papers and their ratio in the skeleton and remainder. The distribution of paper points looks similar to computer science as well, albeit with larger absolute numbers. Regarding the age of the collaborating researchers, the most scholars were born after 1990 in the skeleton and between 1985 and 1990 in the remainder. Thus, there are younger scientists in the convex skeleton of the physics network than in the remainder, which is also different from computer science. So it would appear that low strength ties in physics (most of them being in the remainder) are collaborations between more senior scholars, compared to collaborations between more junior researchers in computer science. Therefore, we may speculate that innovation and interdisciplinary knowledge transfer (see Section 5.5) happens by junior scholars in computer science and by more senior ones in physics. The distribution of the difference in age has a similar shape like that in computer science but with a more pronounced distinction between the skeleton, which has higher values, and the remainder. The interpretation of the last two charts in the second row of plots in Figure 7 is as follows: the share of co-authorships between authors of an equal status and of the same gender in the total number of collaborations is much bigger in the skeleton then in the remainder, which is in a stark contrast with the situation in computer science where the relations between the skeleton and the remainder are approximately the same.

Insert Figure 7 here.

As far as the distribution of the number of co-authors' papers is concerned, the shapes of the skeleton and remainder curves look somewhat similar to computer science, with also two peaks for the skeleton (at 100 and after 300) and only one (before 100) for the remainder. A high similarity of the two scientific fields appears also with respect to the difference in the number of papers. But, again, there is a clearer distinction between the skeleton and the remainder in physics, with higher values for the former. This would mean that collaborations between authors with a rather large difference in scientific production tend to be retained in the physics convex skeleton. Given the similar layout of the skeleton and remainder curves with respect to the age difference mentioned above, we can make a reasonable guess that the less productive and more productive researchers are (doctoral) students and their advisors, respectively. As regards the collaborations between different scientific fields, their share in physics is smaller than in computer science, making physics a less interdisciplinary field. Interdisciplinarity in the skeleton appears to be even smaller than in the remainder. Most collaborators have a PhD degree, which is only slightly more pronounced than in computer science. Compared to computer science, physics papers generally receive more points which relates to their being mostly journal articles and quite frequently "prime papers". The most collaborations take place between researchers whose total sum of points is slightly lower than 2000 in the skeleton as well as in the remainder. The distribution of points difference follows the same line like in computer science with larger absolute values and a more visible superiority of the skeleton over the remainder. On the other hand, the layout of co-authors' prime paper points is completely different: there is a first peak at 0.5 in both the skeleton and the remainder and a second one at 0.9 in the skeleton only. That means that there are many collaborations in the skeleton between researchers with prime papers only. The difference in prime papers, as can be seen in the bottom-right chart in Figure 4, declines more smoothly than in computer science with higher values in the skeleton. Still, the most collaborating authors have no prime papers difference at all.

Table S.2 in the electronic supplement shows the top 20 physicists ranked by four centrality measures. 15 of them are uniquely top-ranked by one of the indicators based on the original co-authorship network and 24 based on the convex skeleton of that network. Robert Blinc [4], Dragan Mihailović [4540], and Janez Dolinšek [3939] regularly appear among the top researchers in the network as well as in the skeleton. They are undoubtedly very influential Slovenian physicists. On the other hand, there are other scholars like Zvonko Trontelj [208] and Cene Filipič [4347] that appear among the top scientists in the skeleton (by any measure except for betweenness centrality) but not in the network. Here we may suggest as well that their prominent position in the convex skeleton is due to their ability of reaching out to rather isolated physicists or to some scientists in distant and rare subfields of physics. It should also be noted that the current (2017) most frequently cited Slovenian physicist according to Web of Science, Matjaž Perc, is completely absent from Table S.2. This is certainly due to the fact that he has collaborated mostly with researchers outside of Slovenia in the period under study. The structure of the whole physics network is presented in Figure S.8 in the electronic supplement.

*5.3. Sociology*

Figure 8 depicts various data distributions in the convex skeleton and the remainder of the co-authorship network of sociologists. The most collaborations took place around 2000 in the skeleton but a few years later (in 2005 and afterwards) in the remainder, which is somewhat similar to computer science. However, the most frequent paper types are neither conference papers like in computer science nor journal articles like in physics, but documents called "Complete Scientific Database or Corpus" (a special category present only in sociology) of which there are many more in the skeleton than in the remainder. Because of this publication structure, most sociology papers are not prime papers, though, and the most co-authored publications received no points at all. This fact is even more pronounced in the skeleton than in the remainder (see the top-right chart in Figure 8). The distribution of co-authorships by the average birth year of co-authors in the skeleton is clearly different from that in the remainder. In the skeleton there is only one major peak at about 1965 whereas in the remainder there is quite a flat evolution between 1970 and 1990. The skeleton thus seems to highlight the collaboration of more senior researchers (see the first plot in the second row of Figure 8), which is consistent with computer science but distant from physics. The most frequent difference in the age of collaborators is zero years in the skeleton, but not so in the remainder where a difference of about seven years appears more often. More collaborations occur between persons with the same status than between those with a different one (e.g. an advisor and a student) and this relation is roughly the same in the skeleton like in the remainder. What makes sociology differ from computer science and physics is the proportion of inter-gender co-authorships, which is significantly higher here than in those two disciplines. However, it is about the same (around 50% of inter-gender ties) in the skeleton as in the remainder.

Insert Figure 8 here.

The most collaborations occurred between authors with approximately 70 papers in both the skeleton and the remainder (see the first plot in the third row of charts in Figure 5). However, there is a secondary "post-peak" in the skeleton with nearly 150 papers and a secondary "pre-peak" in the remainder with about 30 papers. The distribution of co-authorships by the difference in the number of co-authors' papers follows a similar curve like in computer science and physics with visibly more collaborations present in the skeleton than in the remainder that have a difference larger than 100. The proportion of inter-field ties to intra-field ties is not depicted in Figure 8 as there are no inter-field ties. On the other hand, the proportion of collaborations between authors of whom one has no PhD degree is clearly higher than in computer science and physics and is even more pronounced in the skeleton than in the remainder. As regards the distribution of collaborations by the number of the research evaluation points obtained by co-authors, the bottom-left chart depicts a primary peak at about 500 and a secondary one at around 1500 in the skeleton and the whole structure lowered and somewhat shifted to the right with peaks before 1000 and after 1500 in the remainder. The data points in the next chart representing the number of ties between researchers with varying differences in the number of points achieved in the Slovenian research evaluation system are quite scattered. The point difference of about 2000 is markedly more present in the skeleton than in the remainder. Due to the nature of publication types in sociology, there are almost no prime papers and most collaboration thus take place between authors with prime points close to zero (see bottom-right charts in Figure 8).

  The different nature of the collaboration network in sociology when compared to computer science or physics is also reflected in the author rankings in Table S.3 in the electronic supplement. Here the numbers of uniquely ranked researchers in the network and in the skeleton are quite close: 8 and 11, respectively. The always top-ranked sociologists are Drago Kos [9735] and Niko Toš [2469] who are both highly esteemed in both the network and the skeleton, irrespective of the indicator used. Ivan Svetlik [4244] appears among the top scientists everywhere except the convex skeleton researchers ranked by closeness. Zdenko Roter [387] and Pavel Gantar [3604] are hot candidates for "collaboration hubs" too because their presence among the top skeleton authors and absence from the top network authors indicates that they may connect remote and self-contained areas of sociology.

*5.4. Centrality correlations*

The correlation coefficients Spearman's rho ($\rho$) and Kendall's tau ($\tau$) between the rankings by four centrality measures generated from the networks and from the convex skeletons are shown in Figures 9, 10, and 11 (left-hand charts). Following the diagonal entries in the tables, we can get a glimpse how well a convex skeleton ranking correlates to the original network ranking using a particular metric for the nodes. Basically, the convex skeleton rankings are quite well correlated ($\rho \approx 0.8$, $\tau \approx 0.6$), which is in a stark contrast to the small correlation of the rankings produced from the maximum spanning trees ($\rho \approx 0.4$, $\tau \approx 0.3$) shown in the second column of charts. In addition to convex skeletons and maximum spanning trees we also computed correlations of rankings based on the backbones consisting of the same number of high-betweenness or high-embeddedness edges as in the convex skeletons (the third and fourth column of charts in Figures 9, 10, and 11). There it appears that the former correlates slightly worse than the convex skeleton except for betweenness and closeness ($\rho \approx 0.7$, $\tau \approx 0.6$) and the latter generally correlates almost as well as the convex skeleton ($\rho \approx 0.7$, $\tau \approx 0.6$).

Insert Figure 9 here.

Insert Figure 10 here.

Insert Figure 11 here.

*5.5. Possible applications*

The results presented in Sections 5.1 – 5.3 were all based on the fractional counting scheme of collaborations. We also tested the other counting techniques (full and partial) and show the frequency data distributions at least for physics in the appendix (Figure A.1 and Figure A.2). The distributions are very much the same as in Figure 7. Also in the appendix, there is Figure A.3 depicting the results for mathematics and Figure A.4 for economics along with the corresponding correlation charts in Figure A.5 and Figure A.6. (The results yielded by all three counting techniques in all five scientific fields along with the top 20 researchers by all four centrality measures can be found in the electronic supplement.) Nevertheless, we believe that the low strength links that need to be removed in the process of identifying a convex skeleton may, in fact, bear more information than the convex skeleton itself. Weak links (or ties) were defined by Granovetter (1973) as edges connecting mostly remote strongly-tied communities in a social network. Without such ties a flow of information among different communities of a social network would be impossible or reduced (Onnela et al., 2007).

However, weak links in our data usually have a low co-authorship weight as can be seen in Tables 1, 2, and 3 and it seems that the convexity algorithm basically removes some weak ties from the original network. By definition, weak links in our data are those with low co-authorship weight and it seems that the convexity algorithm basically removes some weak ties from the original network. By Granovetter's hypothesis, these are usually not embedded in dense parts and have a low clustering coefficient, while they usually connect distinct parts of the network. Our removal procedure tries to increase the overall network clustering as much as possible. Therefore, it probably removes ties that connect distinct parts of the network that presumably have low strengths. Therefore, we do not intentionally remove weak ties and there actually might be even weaker ties in the network that remain. but the fact that removed ties have low strengths is a consequence of the removal of edges in the extraction of a convex skeleton.

Even though in the broadest sense the identification of a convex skeleton in a network is actually a graph-reduction technique and could be used as such in general, one of the possible applications is surely thinkable in the context of scientometrics and science policy. Scientometricians could apply the concept of convex skeletons to a collaboration network of researchers in a certain country, field, or institution in order to detect the weak ties between scholars. If the detected weak ties are found unnecessary, science policy makers could then adjust the allocation of research funding to discourage the collaboration of scientists leading to such weak ties. This would happen without the need to change the structure of research because in a convex skeleton, as we pointed out, all nodes and strong links are retained. Thus, the research performance of an already well-performing system could be further improved by this approach.

Another feasible application of convex skeletons is using them as a part of a collaboration recommendation system. Provided that the shortest paths between any two nodes in a convex skeleton (and there is exactly one such shortest path between any two nodes there if the network is fully convex) have, on average, larger weights than the shortest paths between any two nodes in the original network, they would represent the recommended paths from one researcher to another, optimized for the maximum weights of the nodes traversed. These weights might, for instance, be some centrality measures (Abbasi et al., 2012) or numbers of points obtained by scientists in a research evaluation system. Thus, choosing potential future collaborators on the basis of a convex skeleton in a co-authorship network already in place may be a good strategy to maintain or improve one's own position (Abbasi et al., 2011) in the increasingly competitive scientific community.

## 6. Conclusions and future work

Network convexity is a property of undirected connected graphs that can be measured and that always falls within the interval [0, 1]. A fully convex graph has a convexity of 1 and only consists of subgraphs such that each of them includes all shortest paths between all its nodes. A convex skeleton is the largest fully convex component of a graph that is created by removing the least number of edges from the original graph. It is a backbone network, but, unlike a spanning tree, it does contain cycles within cliques. In general terms, it is a tree of cliques and has some advantages over a spanning network because it is a reduced graph that retains the most important structural properties of the original network. The goal of this study was to apply the notion of convexity and convex skeletons in the context of scientific collaboration (co-authorship) networks and to find out whether there could be some possible applications in scientometrics.

For this purpose, we took the following steps:

- We analyzed datasets of Slovenian researchers' collaborations in computer science, physics, and sociology.
- We determined the convex skeletons in those collaboration networks by removing low strength links from them.
- We generated frequency distributions of various data parameters of the skeletons and of the residual graphs and compared them thoroughly in order to find structural differences.
- We also calculated four different centrality indicators for the scholars in both the original co-authorship networks and the convex skeletons and juxtaposed the top 20 rankings
- We verified our approach also in the context of other scientific fields: mathematics and economics.

A major contribution of this study is the development of a technique to identify influential researchers that were previously "hidden" in the collaboration network of their scientific community by applying standard centrality computations to the convex skeleton extracted from that collaboration network. Based on our experiments we found that:

- Scholarly collaboration networks are rather convex (see Convexity in Tables 1, 2, and 3), which means that their topological arrangement is somewhat close to a tree of cliques instead of a spanning tree or a random graph (see Fig. 1).
- The feature of convex skeletons to highlight older collaborations and collaborations between more senior researchers is more pronounced in computer science and sociology than in physics.
- There are potentially very influential scientists who are not prominent in the collaboration networks of their scholarly communities.
- Convex skeletons are a good abstraction of the original co-authorship networks because quite often the same authors appear at the top of the rankings by various centrality measures based on the original network and the skeleton.

The potential of convexity and convex networks is by far not exhausted by the present study. In addition to further extended analyses including also other scientific disciplines or countries, other possible applications of convex skeletons in scientometrics and science policy need to be investigated in our future work. As already discussed, the collaboration edges whose removal leads to the creation of a convex skeleton can be considered as weak and having no impact on the structure of the whole collaboration network. They may thus be found unnecessary and science policy makers could easily reduce funding for the projects that generate these ties. We assume that, in the long term, this could lead to more efficient research policy systems. Also, a significant limitation of this study is the "locality" of collaboration because interdisciplinary research is not included. If it were, this would likely have an effect on the global convexity measure of the collaboration network and the usefulness of convex skeletons in the detection of weak ties would probably be even more pronounced. But to verify this assumption, some further research is needed.

**Acknowledgements**

This work has been supported in part by the Slovenian Research Agency under the programs P2-0359 and P5-0168, and by the European Union COST Action number CA15109. It was also supported in part by the Ministry of Education, Youth and Sports of the Czech Republic under grant no. LO1506. Furthermore, we would like to thank the anonymous reviewers and the editor for their insightful comments that helped us improve the paper.

**References**

Abbasi, A., Altmann, J., & Hossain, L. (2011). Identifying the effects of co-authorship networks on the performance of scholars: A correlation and regression analysis of performance measures and social network analysis measures. *Journal of Informetrics*, 5(4), 594-607.

Abbasi, A., Hossain, L., & Leydesdorff, L. (2012). Betweenness centrality as a driver of preferential attachment in the evolution of research collaboration networks. *Journal of Informetrics*, 6(3), 403-412.

Adai, A. T., Date, S. V., Wieland, S., & Marcotte, E. M. (2004). LGL: Creating a map of protein function with an algorithm for visualizing very large biological networks. *Journal of Molecular Biology*, 340(1), 179-190.

Barabási, A. L., Jeong, H., Néda, Z., Ravasz, E., Schubert, A., & Vicsek, T. (2002). Evolution of the social network of scientific collaborations. *Physica A: Statistical Mechanics and its Applications*, 311(3-4), 590-614.

Barbosa, M. W., Ladeira, M. B., & de la Calle Vicente, A. (2017). An analysis of international coauthorship networks in the supply chain analytics research area. *Scientometrics*, 111(3), 1703-1731.

Bozeman, B., & Boardman, C. (2014). Research Collaboration and Team Science: A State-of-the-Art Review and Agenda. Springer International Publishing.

Brandão, M. A., & Moro, M. M. (2017). The strength of co-authorship ties through different topological properties. *Journal of the Brazilian Computer Society*, 23(1).

Ciglarič, T. (2017). Convex skeleton. https://github.com/t4c1/Convex-Skeleton.

Dehdarirad, T., & Nasini, S. (2017). Research impact in co-authorship networks: A two-mode analysis. *Journal of Informetrics*, 11(2), 371-388.

Ding, Y. (2011). Scientific collaboration and endorsement: Network analysis of coauthorship and citation networks. *Journal of Informetrics*, 5(1), 187-203.

Farber, M., & Jamison, R. (1986). Convexity in graphs and hypergraphs. *SIAM Journal on Algebraic and Discrete Methods*, 7(3), 433–444.

Ferligoj, A., Kronegger, L., Mali, F., Snijders, T. A. B., & Doreian, P. (2015). Scientific collaboration dynamics in a national scientific system. *Scientometrics*, 104(3), 985-1012.

Fiala, D. (2013). From CiteSeer to CiteSeerX: Author rankings based on coauthorship networks. *Journal of Theoretical and Applied Information Technology*, 58(1), 191-204.

Franceschet, M. (2011). Collaboration in computer science: A network science approach. *Journal of the American Society for Information Science and Technology*, 62(10), 1992-2012.

Glänzel, W. (2001). National characteristics in international scientific co-authorship relations. *Scientometrics*, 51(1), 69-115.

Granovetter, M. S. (1973). The Strength of Weak Ties. *American Journal of Sociology*, 78(6), 1360-1380.

Hara, N., Solomon, P., Kim, S.-L., Sonnenwald, D.H. (2003). An emerging view of scientific collaboration: Scientists' perspectives on collaboration and factors that impact collaboration. *Journal of the American Society for Information Science and Technology*, 54(10), 952-965.

Harary, F., & Nieminen, J. (1981). Convexity in graphs. *Journal of Differential Geometry*, 16(2), 185–190.

Hsu, J., & Huang, D. (2011). Correlation between impact and collaboration. *Scientometrics*, 86(2), 317-324.

Huang, J., Zhuang, Z., Li, J., & Giles, C. L. (2008). Collaboration over time: Characterizing and modeling network evolution. In Proceedings of the 2008 International Conference on Web Search and Data Mining, pp. 107-116, Palo Alto, USA.

Katz, J. S. (1994). Geographical proximity and scientific collaboration. *Scientometrics*, 31(1), 31-43.

Katz, J. S., & Martin, B. R. (1997). What is research collaboration? *Research Policy*, 26(1), 1-18.

Kim, J., & Diesner, J. (2015). The effect of data pre-processing on understanding the evolution of collaboration networks. *Journal of Informetrics*, 9(1), 226-236.

Kim, J., & Diesner, J. (2016). Distortive effects of initial-based name disambiguation on measurements of large-scale coauthorship networks. *Journal of the Association for Information Science and Technology*, 67(6), 1446-1461.

Kim, J., & Diesner, J. (2017). Over-time measurement of triadic closure in coauthorship networks. *Social Network Analysis and Mining*, 7(1).

Kretschmer, H. (2004). Author productivity and geodesic distance in bibliographic co-authorship networks, and visibility on the Web. *Scientometrics*, 60(3), 409-420.

Krumov, L., Fretter, C., Müller-Hannemann, M., Weihe, K., & Hütt, M.-T. (2011). Motifs in co-authorship networks and their relation to the impact of scientific publications. *The European Physical Journal B*, 84(4), 535–540.

Kumar, S., & Jan, J. M. (2014). Research collaboration networks of two OIC nations: comparative study between Turkey and Malaysia in the field of 'Energy Fuels', 2009–2011. *Scientometrics*, 98(1), 387–414.

Laudel, G. (2002). What do we measure by co-authorships? *Research Evaluation*, 11(1), 3-15.

Leifeld, P., Wankmüller, S., Berger, V. T. Z., Ingold, K., & Steiner, C. (2017). Collaboration patterns in the German political science co-authorship network. *PLoS ONE*, 12(4).

Leydesdorff, L., & Park, H. W. (2017). Full and fractional counting in bibliometric networks. *Journal of Informetrics*, 11(1), 117–120.

Liu, X., Bollen, J., Nelson, M. L., & Van De Sompel, H. (2005). Co-authorship networks in the digital library research community. *Information Processing and Management*, 41(6), 1462-1480.

Mali, F., Kronegger, L., Doreian, P., & Ferligoj A. (2012). Dynamic Scientific Co-Authorship Networks. In Scharnhorst, A., Börner, K., & van den Besselaar P. (Eds.) *Models of Science Dynamics. Understanding Complex Systems* (pp. 195-232), Heidelberg: Springer Verlag.

Marc, T., & Šubelj, L. (2018). Convexity in complex networks. *Network Science*, 6(2), 176-203.

Melin, G., & Persson, O. (1996). Studying research collaboration using co-authorships. *Scientometrics*, 36(3), 363-377.

Newman, M. E. J. (2001a). Scientific collaboration networks. I. network construction and fundamental results. *Physical Review E - Statistical, Nonlinear, and Soft Matter Physics*, 64(1 II), art. no. 016131.

Newman, M. E. J. (2001b). Scientific collaboration networks. II. shortest paths, weighted networks, and centrality. *Physical Review E - Statistical, Nonlinear, and Soft Matter Physics*, 64(1 II), art. no. 016132.

Newman, M. E. J. (2001c). The structure of scientific collaboration networks. *Proceedings of the National Academy of Sciences of the United States of America*, 98(2), 404-409.

Newman, M. E. J. (2004). Coauthorship networks and patterns of scientific collaboration. *Proceedings of the National Academy of Sciences of the United States of America*, 101(SUPPL. 1), 5200-5205.

Newman, M. E. J. (2010). Networks: An Introduction. Oxford University Press.

Newman, M. E. J., Strogatz, S. H., & Watts, D. J. (2001). Random graphs with arbitrary degree distributions and their applications. *Physical Review E*, 64(2), art. no. 026118.

Onnela, J.-P., Saramäki, J., Hyvönen, J., Szabó, G., Lazer, D., Kaski, K., Kertész, J., & Barabási, A.-L. (2007). Structure and tie strengths in mobile communication networks.

*Proceedings of the National Academy of Sciences of the United States of America*, 104(18), 7332-7336.

Perc, M. (2010). Growth and structure of Slovenia's scientific collaboration network. *Journal of Informetrics*, 4(4), 475-482.

Perianes-Rodriguez, A., Waltman, L., & van Eck, N. J. (2016). Constructing bibliometric networks: A comparison between full and fractional counting. *Journal of Informetrics*, 10(4), 1178–1195.

Rodela, R. (2016). On the use of databases about research performance: Comments on Karlovčec and Mladenić (2015) and others using the SICRIS database. *Scientometrics*, 109(3), 2151-2157.

Šubelj, L. (2018). Convex skeletons of complex networks. *Journal of the Royal Society Interface*, 15(145), art. no. 20180422.

Van de Vel, M. L. J. (1993). Theory of convex structures. Amsterdam: North-Holland.

Wagner, C. S., Roessner, J. D., Bobb, K., Klein, J. T., Boyack, K. W., Keyton, J., Rafols, I., & Börner, K. (2011). Approaches to understanding and measuring interdisciplinary scientific research (IDR): A review of the literature. *Journal of Informetrics*, 5(1), 14-26.

Yan, E., & Ding, Y. (2009). Applying centrality measures to impact analysis: A coauthorship network analysis. *Journal of the American Society for Information Science and Technology*, 60(10), 2107-2118.

**Figure captions**

**Fig. 1**  *(left)* Triangular lattice with highlighted convex (green diamonds) and non-convex (grey ellipses) subgraphs. Only bold edges are part of the subgraphs, while dashed nodes and edges show the convex hull of the non-convex subgraph. *(middle)* Fully convex graph, i.e. a tree of cliques, where every connected induced subgraph is convex. The graph consists of a clique on four nodes, 4 cliques on three nodes and 8 cliques on two nodes, i.e. individual edges, which are connected in a tree-like arrangement. *(right)* An example of a non-convex graph, i.e. a random graph, where any non-trivial subgraph is likely non-convex. The meaning of different symbols is consistent between the figures.

**Fig. 2**  Growth of convex subgraphs in the first two steps $t \leq 2$ of the convex expansion algorithm. Notice that non-convex subgraphs in the later steps $t \geq 2$ of the algorithm indicate the absence of an either (locally) tree-like or clique-like structure. The meaning of different symbols is the same as in Fig. 1.

**Fig. 3**  Wiring diagram of a particular realization of a convex skeleton of the computer science network. Edges in the convex skeleton are shown in bold green and the remaining ones in light grey. The sizes of the nodes are proportional to their degrees in the convex skeleton, while the labels are shown only for 20 nodes with the highest degree. The nodes with a clustering coefficient above or equal to 0.5 are shown as green diamonds and the others as grey ellipses. The layout was computed from the original network with the Large Graph Layout (Adai et al., 2004).

**Fig. 4**  Wiring diagram of the convex skeleton of the computer science network shown in Fig. 3. Only edges in the convex skeleton are shown and the layout was computed from the skeleton. Other details are the same as in Fig. 3.

**Fig. 5**  Wiring diagram of a particular realization of a maximum spanning tree of the computer science network. Only edges in the spanning tree are shown and the layout was computed from the tree. Other details are the same as in Fig. 3.

**Fig. 6**  Different distributions for the convex skeletons and the remainder or residual graphs, i.e. graphs with convex skeleton edges removed, extracted from the computer science network. The charts show the results for the fractional counting technique, while the light green symbols or bars represent the skeletons and the dark grey ones represent the remainders.

**Fig. 7**  Different distributions for the convex skeletons and the remainder graphs extracted from the physics network. Other details are the same as in Fig. 6.

**Fig. 8**  Different distributions for the convex skeletons and the remainder graphs extracted from the sociology network. Other details are the same as in Fig. 6.

**Fig. 9**  Correlations between measures of node position in the computer science network (shown in matrix rows) and its backbones (shown in matrix columns). The backbones are the extracted convex skeletons and maximum spanning trees, and the backbones consisting of the same number of high-betweenness or high-embeddedness edges as in the convex skeletons. The measures of node position include node degree and PageRank score, and betweenness and closeness centralities. Top and bottom rows show Spearman's and Kendall's rank correlation coefficients, respectively.

**Fig. 10**  Correlations between measures of node position in the physics network and its backbones. Other details are the same as in Fig. 9.

**Fig. 11**  Correlations between measures of node position in the sociology network and its backbones. Other details are the same as in Fig. 9.

**Fig. A.1**  Different distributions for the convex skeletons and the remainder graphs extracted from the physics network. The charts show the results for the full counting technique, while other details are the same as in Fig. 6.

**Fig. A.2**  Different distributions for the convex skeletons and the remainder graphs extracted from the physics network. The charts show the results for the partial counting technique, while other details are the same as in Fig. 6.

**Fig. A.3**  Different distributions for the convex skeletons and the remainder graphs extracted from the mathematics network. Other details are the same as in Fig. 6.

**Fig. A.4**  Different distributions for the convex skeletons and the remainder graphs extracted from the economics network. Other details are the same as in Fig. 6.

**Fig. A.5**  Correlations between measures of node position in the mathematics network and its backbones. Other details are the same as in Fig. 9.

**Fig. A.6**  Correlations between measures of node position in the economics network and its backbones. Other details are the same as in Fig. 9.

**Table captions**

**Table 1**  Descriptive statistics of the computer science network and particular realizations of its backbones. These show the number of nodes and edges, the fraction of nodes in the largest connected component, the average degree, the average distance between the nodes in the largest connected component, assortativity and clustering coefficients, and corrected network convexity.

**Table 2**  Descriptive statistics of the physics network and particular realizations of its backbones, while other details are the same as in Table 1.

| | |
|---|---|
| **Table 3** | Descriptive statistics of the sociology network and particular realizations of its backbones, while other details are the same as in Table 1. |
| **Table A.1** | Descriptive statistics of the mathematics network and particular realizations of its backbones, while other details are the same as in Table 1. |
| **Table A.2** | Descriptive statistics of the economics network and particular realizations of its backbones, while other details are the same as in Table 1. |

**Appendix**

Insert Figure A.1 here.

Insert Figure A.2 here.

Insert Figure A.3 here.

Insert Figure A.4 here.

Insert Figure A.5 here.

Insert Figure A.6 here.

Insert Table A.1 here.

Insert Table A.2 here.

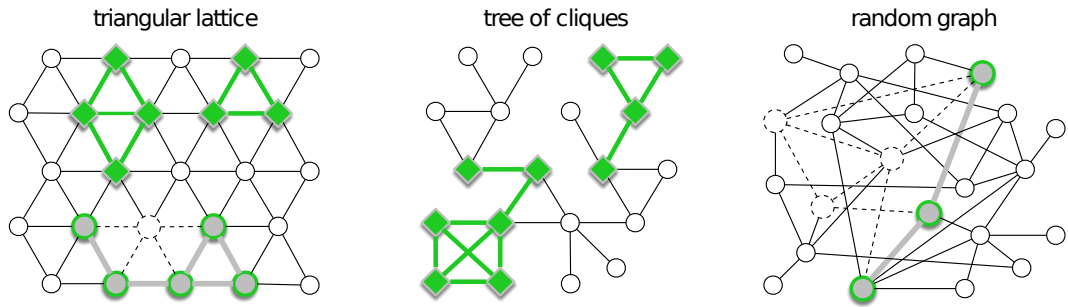

**Fig. 1** *(left)* Triangular lattice with highlighted convex (green diamonds) and non-convex (grey ellipses) subgraphs. Only bold edges are part of the subgraphs, while dashed nodes and edges show the convex hull of the non-convex subgraph. *(middle)* Fully convex graph, i.e. a tree of cliques, where every connected induced subgraph is convex. The graph consists of a clique on four nodes, 4 cliques on three nodes and 8 cliques on two nodes, i.e. individual edges, which are connected in a tree-like arrangement. *(right)* An example of a non-convex graph, i.e. a random graph, where any non-trivial subgraph is likely non-convex. The meaning of different symbols is consistent between the figures.

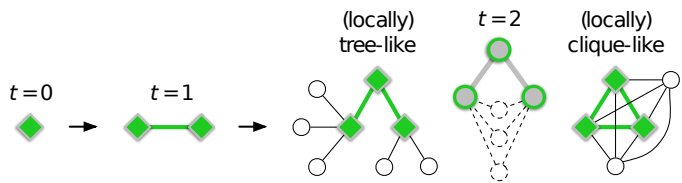

**Fig. 2** Growth of convex subgraphs in the first two steps $t \leq 2$ of the convex expansion algorithm. Notice that non-convex subgraphs in the later steps $t \geq 2$ of the algorithm indicate the absence of an either (locally) tree-like or clique-like structure. The meaning of different symbols is the same as in Fig. 1.

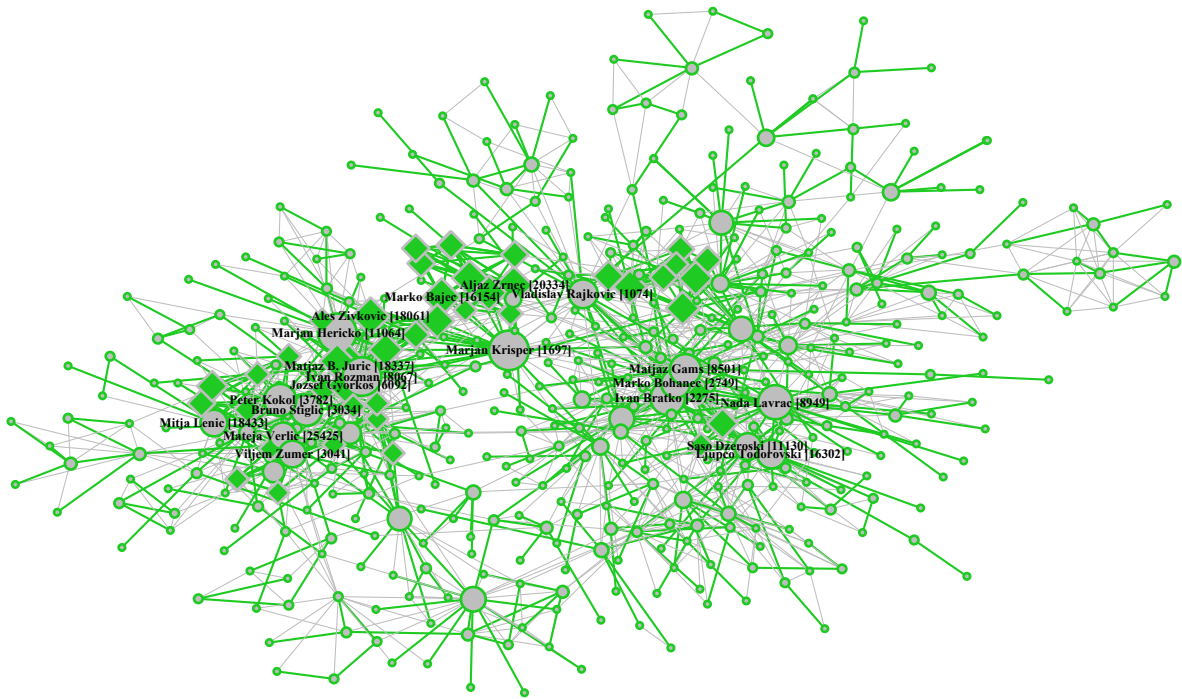

**Fig. 3**  Wiring diagram of a particular realization of a convex skeleton of the computer science network. Edges in the convex skeleton are shown in bold green and the remaining ones in light grey. The sizes of the nodes are proportional to their degrees in the convex skeleton, while the labels are shown only for 20 nodes with the highest degree. The nodes with a clustering coefficient above or equal to 0.5 are shown as green diamonds and the others as grey ellipses. The layout was computed from the original network with the Large Graph Layout (Adai et al., 2004).

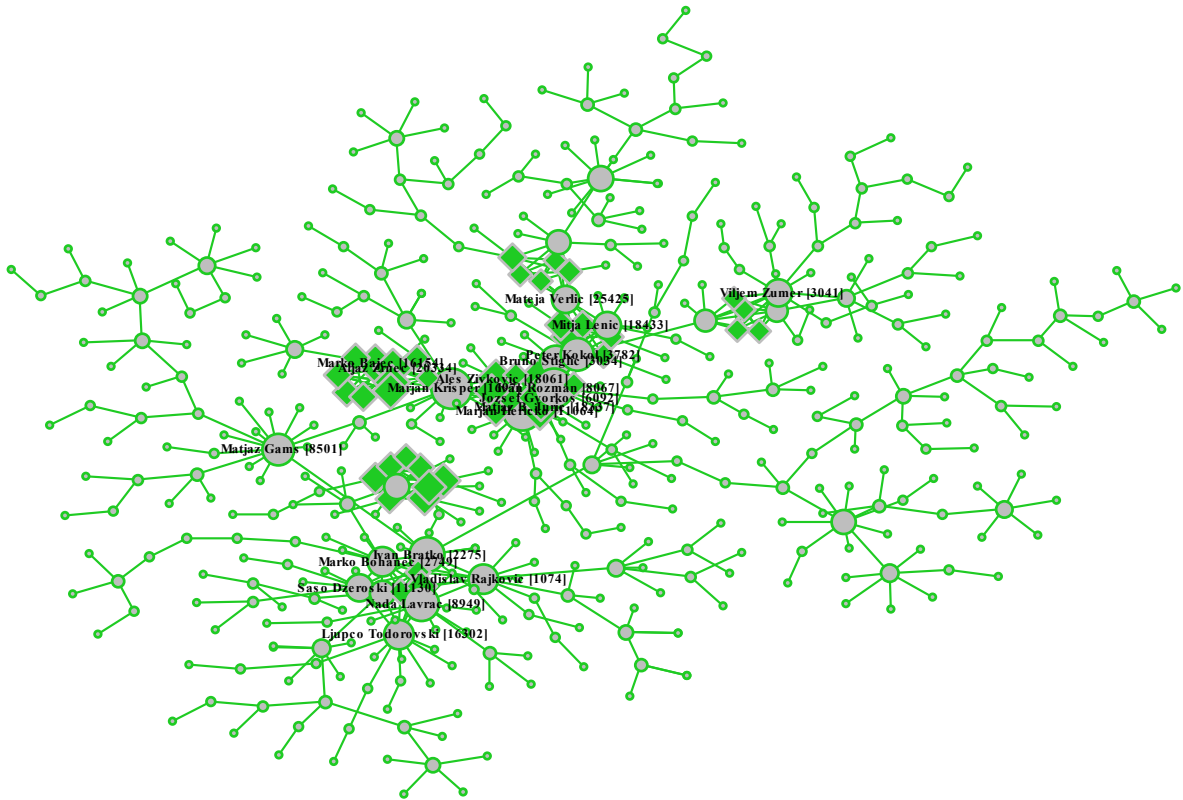

**Fig. 4** Wiring diagram of the convex skeleton of the computer science network shown in Fig. 3. Only edges in the convex skeleton are shown and the layout was computed from the skeleton. Other details are the same as in Fig. 3.

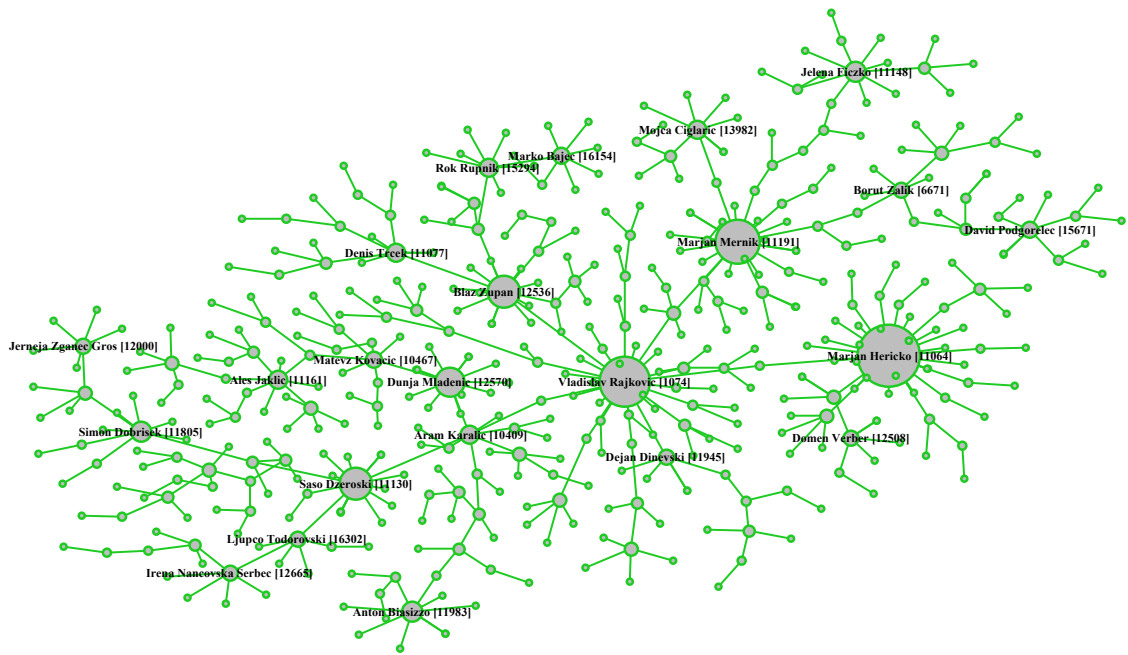

**Fig. 5**    Wiring diagram of a particular realization of a maximum spanning tree of the computer science network. Only edges in the spanning tree are shown and the layout was computed from the tree. Other details are the same as in Fig. 3.

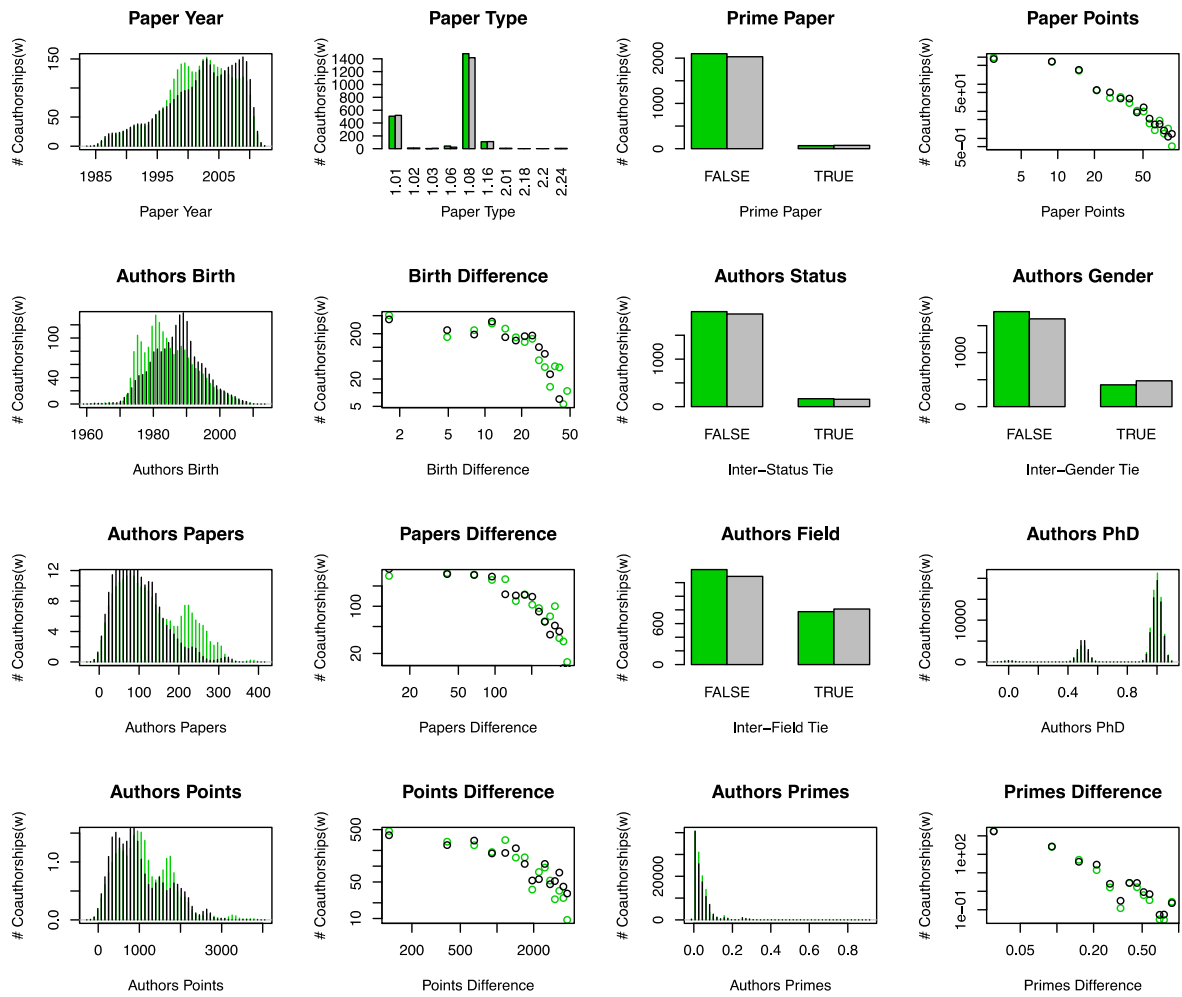

**Fig. 6** Different distributions for the convex skeletons and the remainder or residual graphs, i.e. graphs with convex skeleton edges removed, extracted from the computer science network. The charts show the results for the fractional counting technique, while the light green symbols or bars represent the skeletons and the dark grey ones represent the remainders.

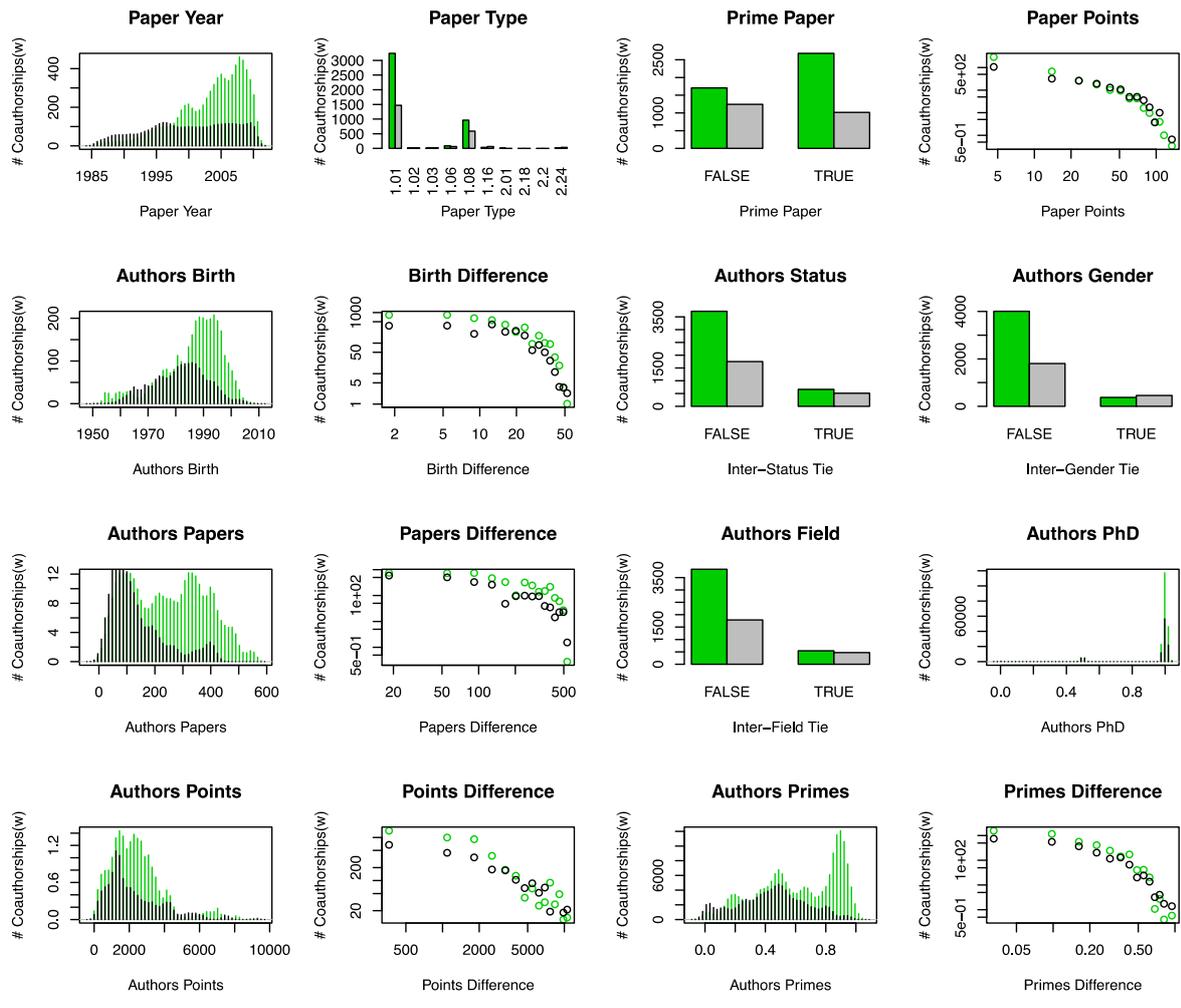

**Fig. 7** Different distributions for the convex skeletons and the remainder graphs extracted from the physics network. Other details are the same as in Fig. 6.

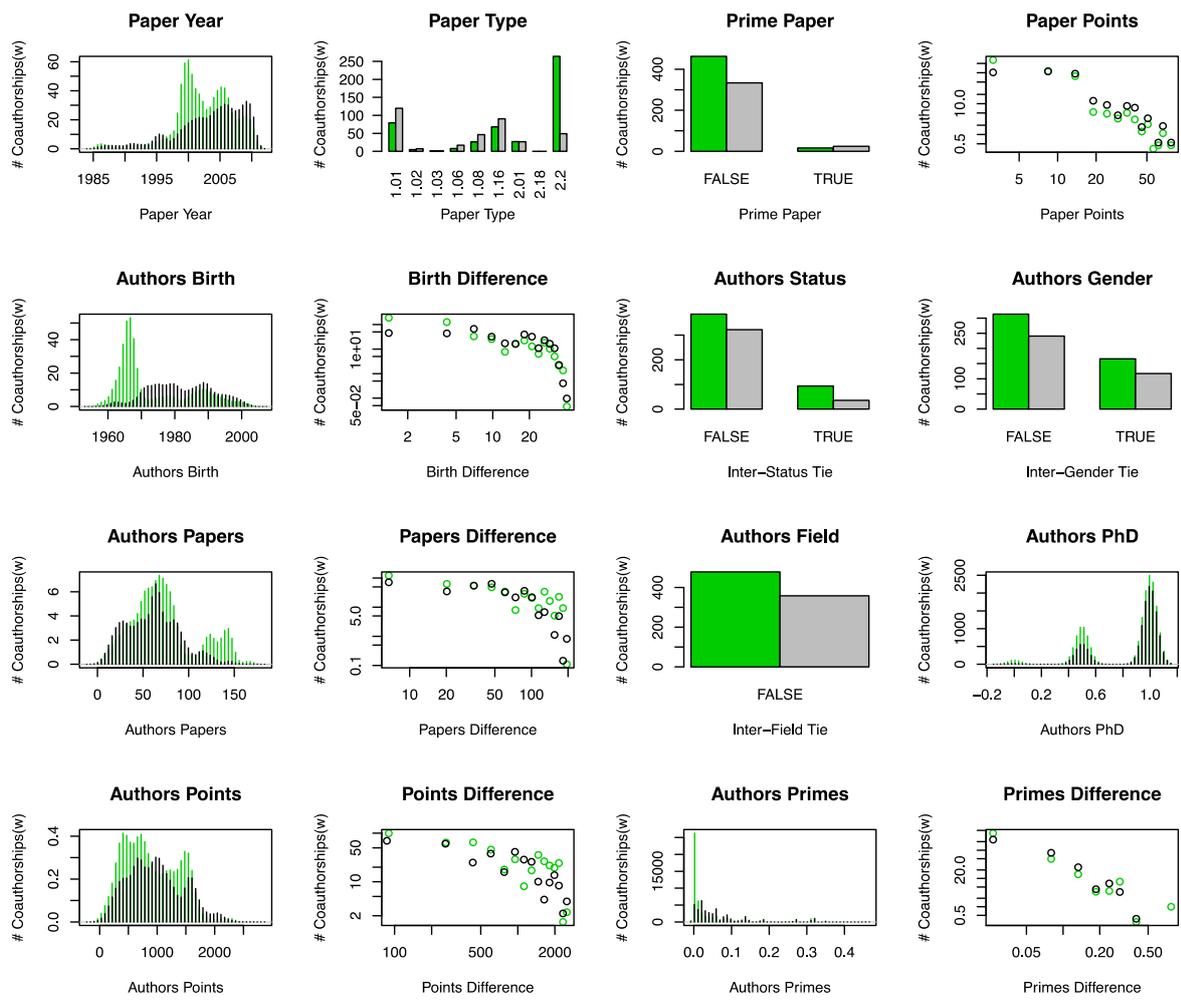

**Fig. 8** Different distributions for the convex skeletons and the remainder graphs extracted from the sociology network. Other details are the same as in Fig. 6.

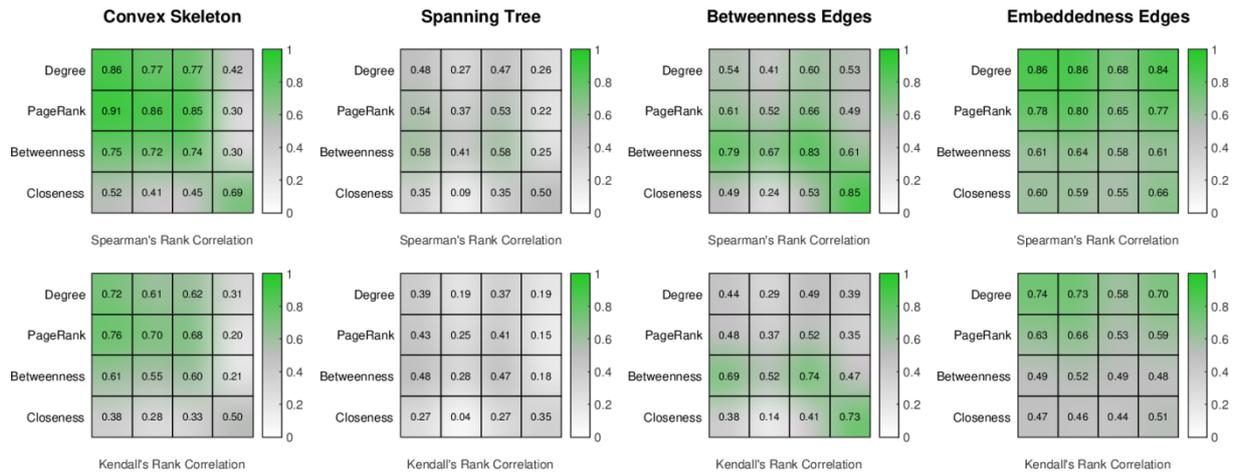

**Fig. 9** Correlations between measures of node position in the computer science network (shown in matrix rows) and its backbones (shown in matrix columns). The backbones are the extracted convex skeletons and maximum spanning trees, and the backbones consisting of the same number of high-betweenness or high-embeddedness edges as in the convex skeletons. The measures of node position include node degree and PageRank score, and betweenness and closeness centralities. Top and bottom rows show Spearman's and Kendall's rank correlation coefficients, respectively.

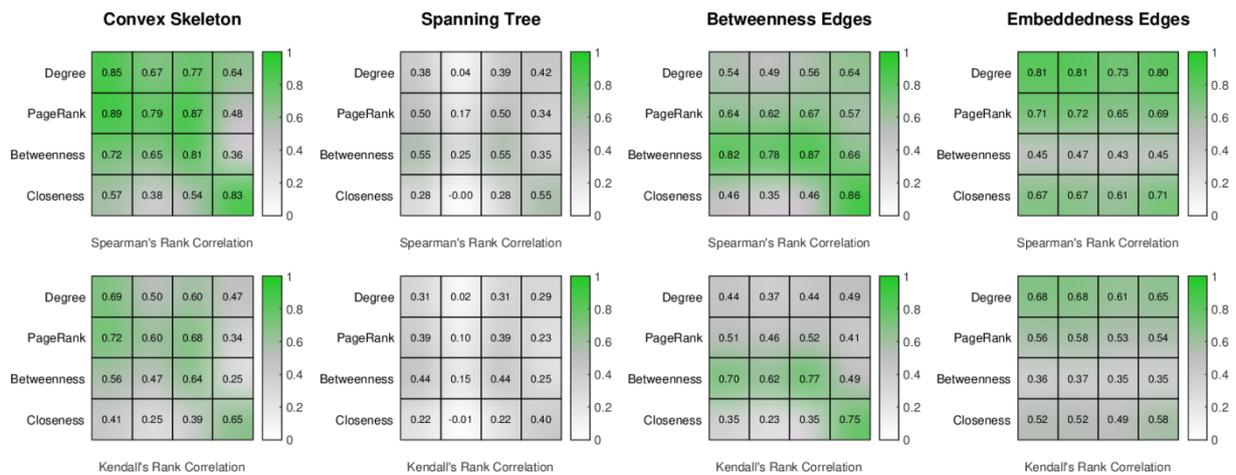

**Fig. 10** Correlations between measures of node position in the physics network and its backbones. Other details are the same as in Fig. 9.

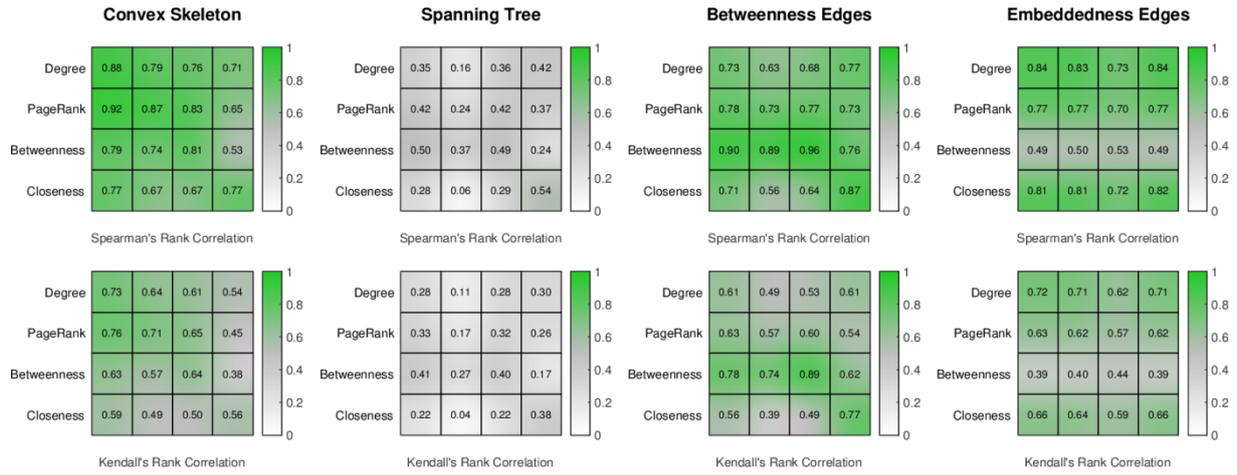

**Fig. 11** Correlations between measures of node position in the sociology network and its backbones. Other details are the same as in Fig. 9.

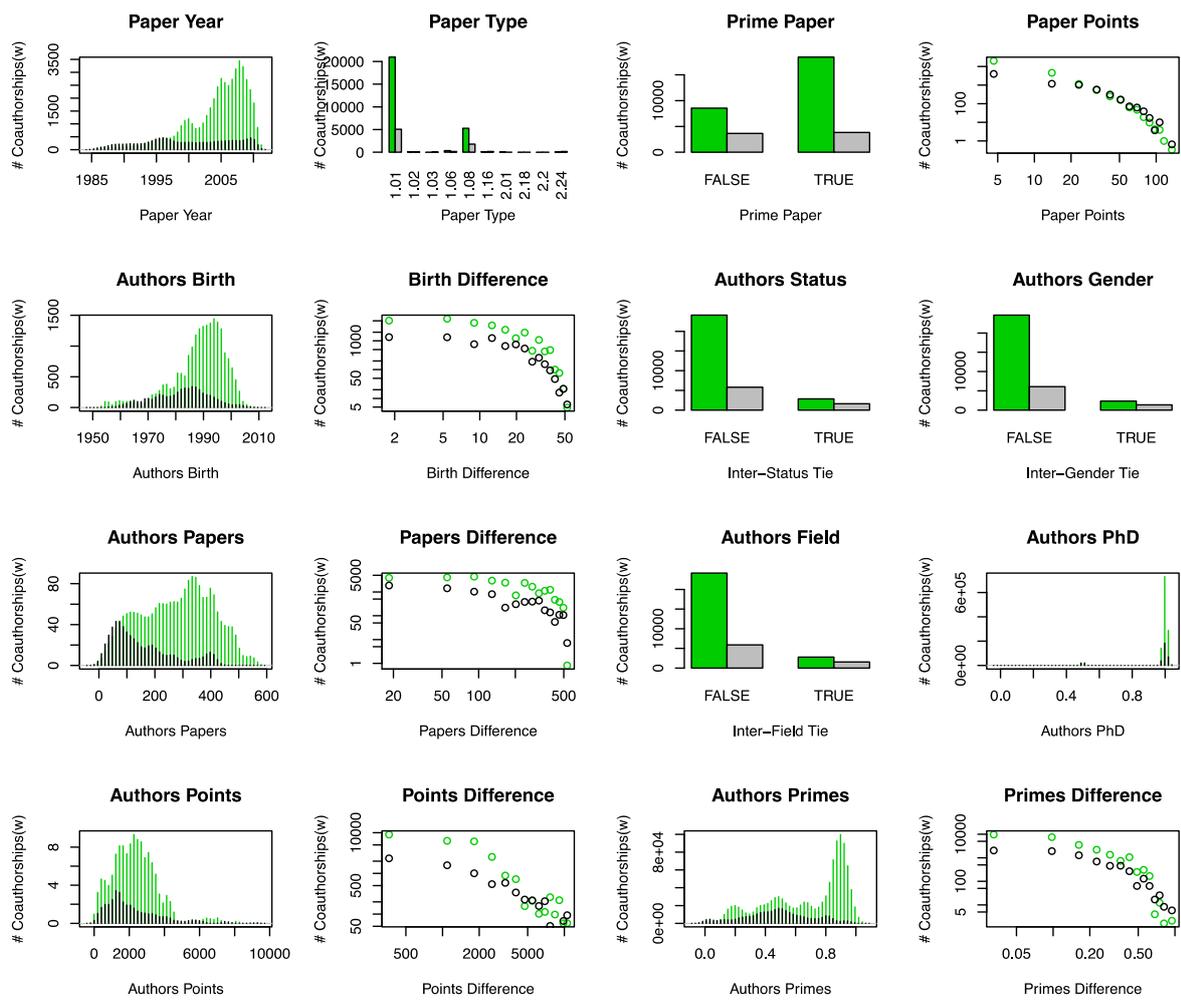

**Fig. A.1** Different distributions for the convex skeletons and the remainder graphs extracted from the physics network. The charts show the results for the full counting technique, while other details are the same as in Fig. 6.

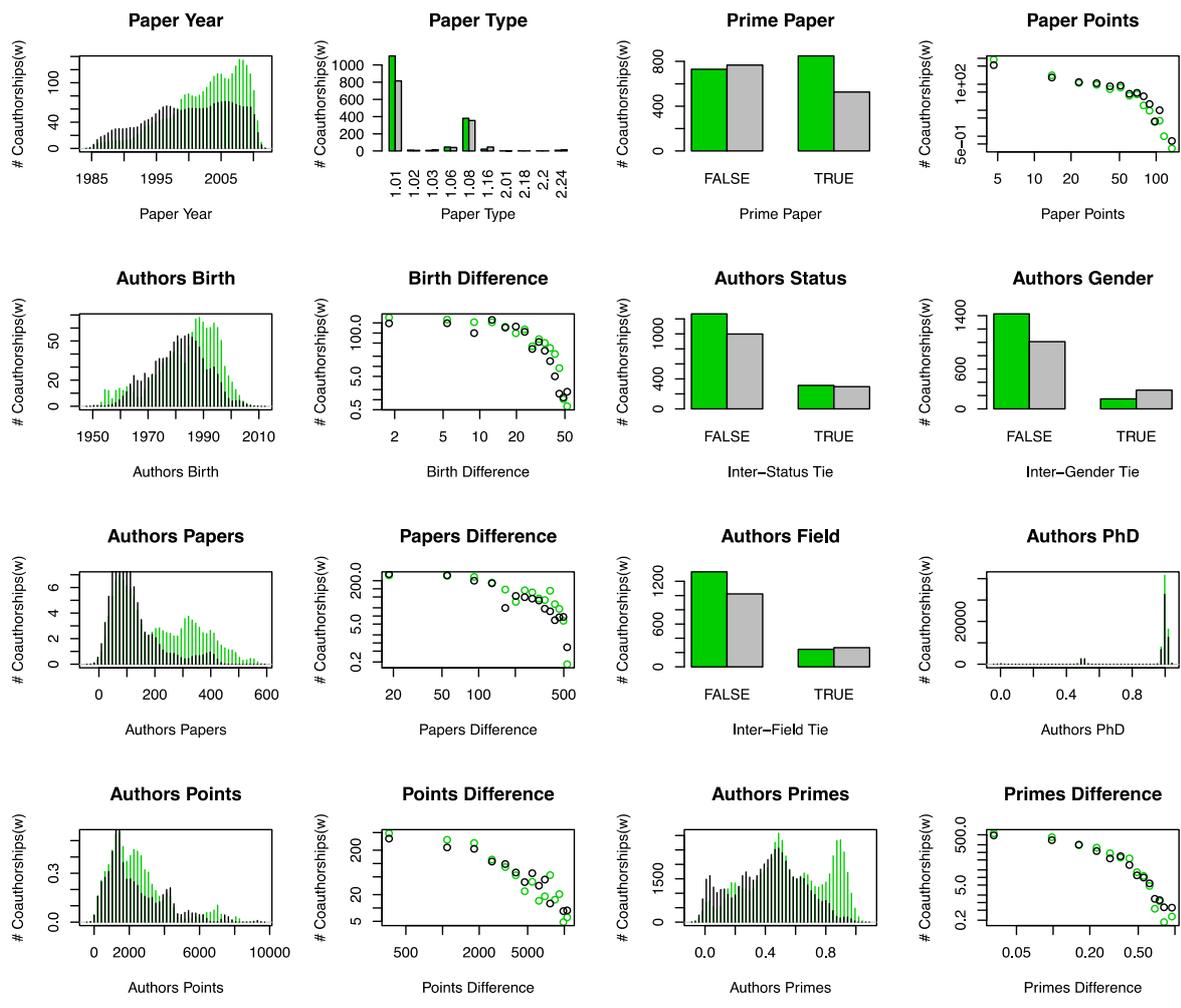

**Fig. A.2** Different distributions for the convex skeletons and the remainder graphs extracted from the physics network. The charts show the results for the partial counting technique, while other details are the same as in Fig. 6.

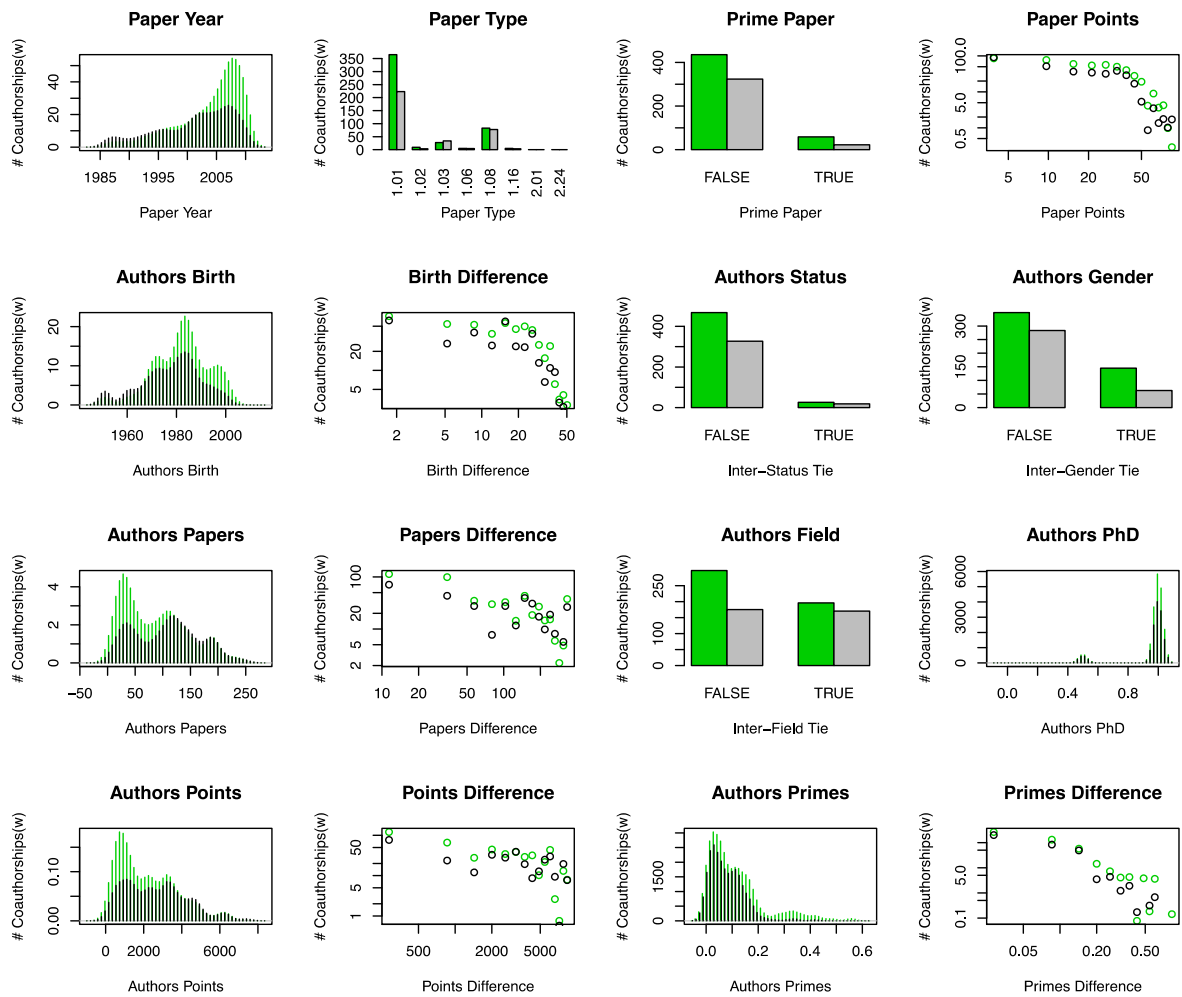

**Fig. A.3** Different distributions for the convex skeletons and the remainder graphs extracted from the mathematics network. Other details are the same as in Fig. 6.

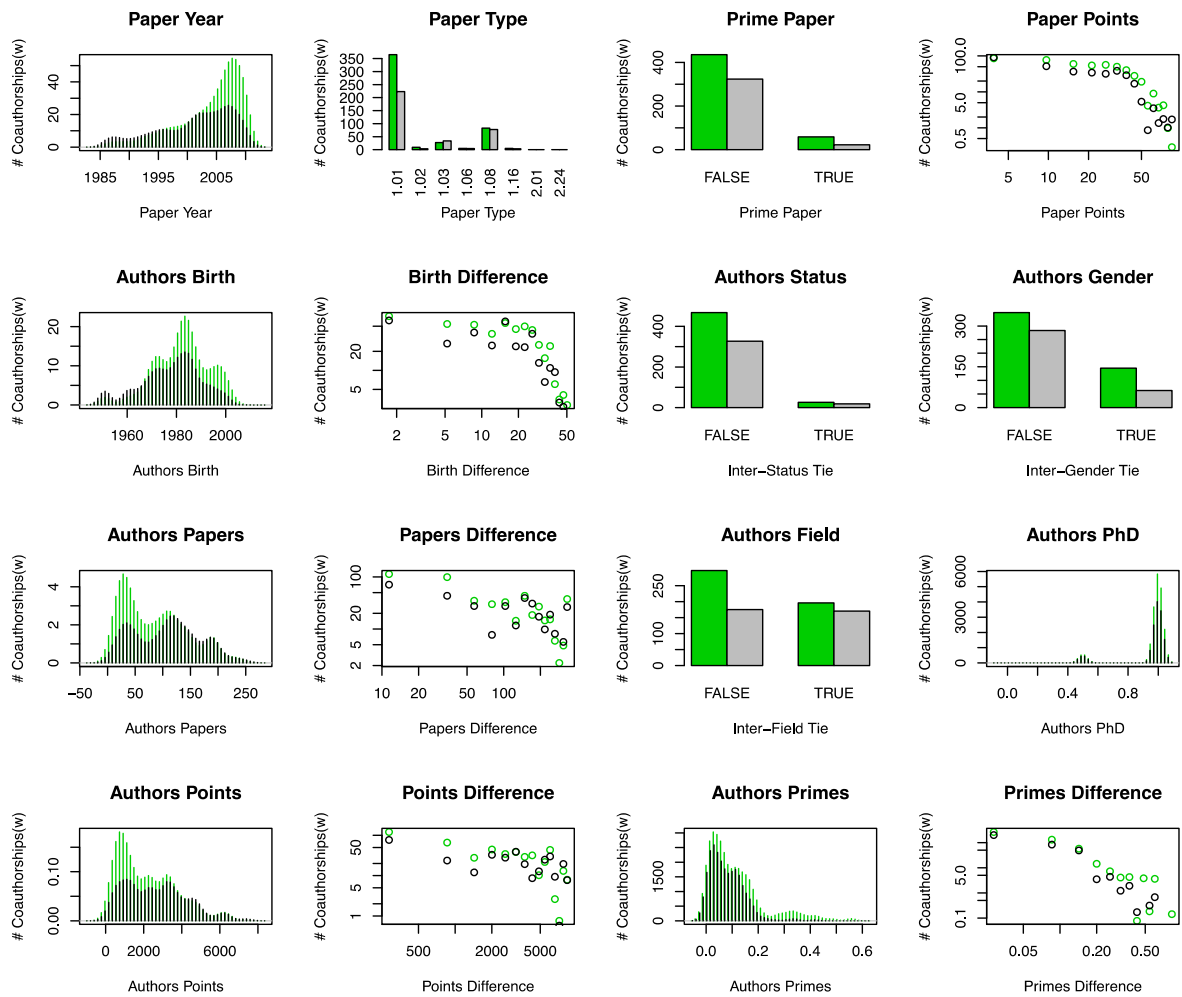

**Fig. A.4** Different distributions for the convex skeletons and the remainder graphs extracted from the economics network. Other details are the same as in Fig. 6.

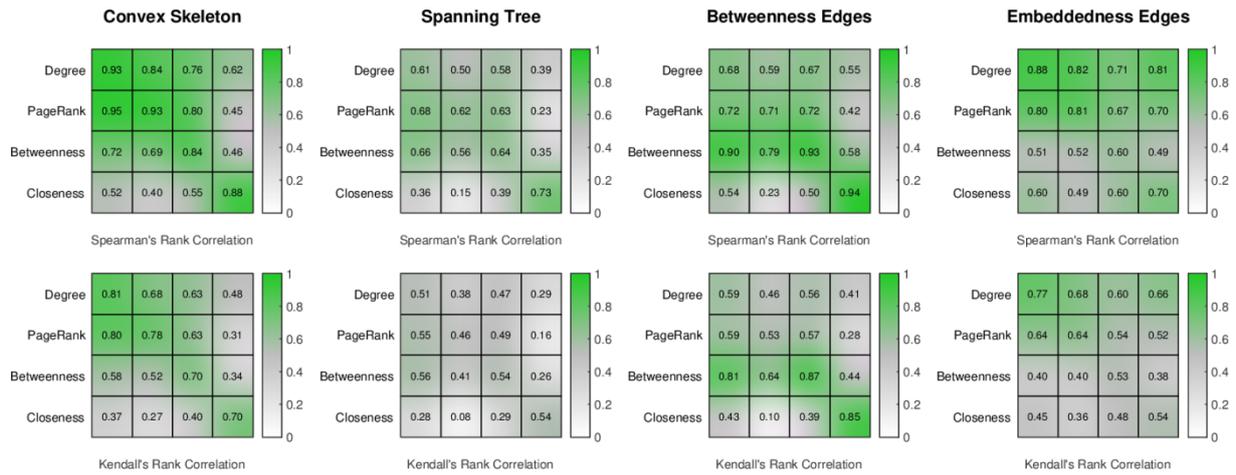

**Fig. A.5** Correlations between measures of node position in the mathematics network and its backbones. Other details are the same as in Fig. 9.

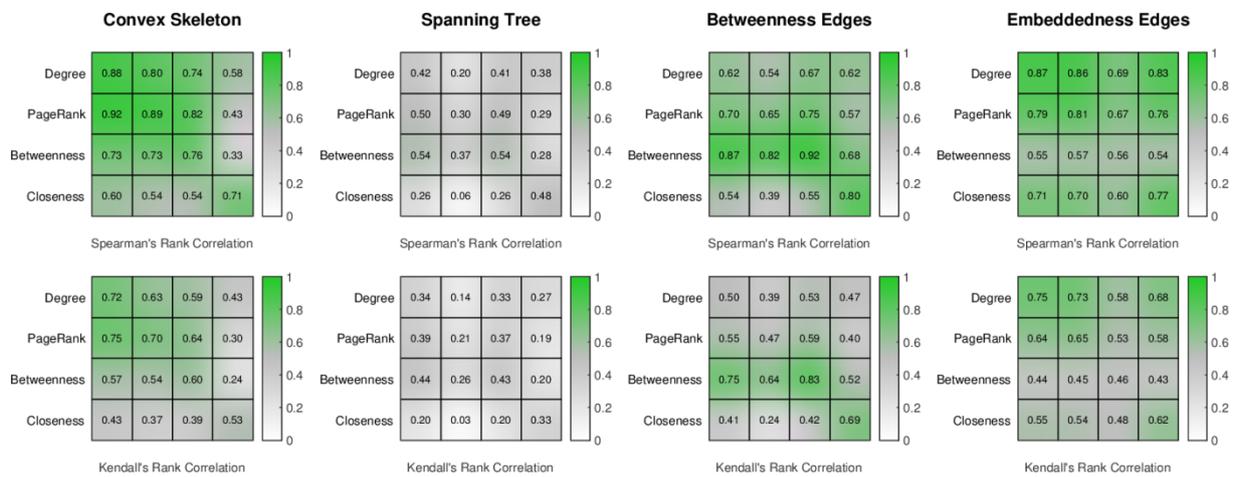

**Fig. A.6** Correlations between measures of node position in the economics network and its backbones. Other details are the same as in Fig. 9.

**Table 1**  Descriptive statistics of the computer science network and particular realizations of its backbones. These show the number of nodes and edges, the fraction of nodes in the largest connected component, the average degree, the average distance between the nodes in the largest connected component, assortativity and clustering coefficients, and corrected network convexity.

|  | Network | Convex Skeleton | Spanning Tree | Betweenness Edges | Embeddedness Edges |
|---|---|---|---|---|---|
| # Nodes | 475 | 475 | 475 | 475 | 475 |
| # Edges | 1,548 | 677 | 474 | 677 | 677 |
| % LCC | 100.0 | 100.0 | 100.0 | 93.7 | 39.6 |
| Degree | 6.52 | 2.85 | 2.00 | 2.85 | 2.85 |
| Distance | 4.11 | 9.55 | 8.55 | 4.26 | 4.41 |
| Assortativity | -0.02 | 0.28 | -0.25 | -0.21 | 0.04 |
| Clustering | 0.55 | 0.09 | 0.00 | 0.06 | 0.28 |
| Convexity | 0.47 | 0.97 | 1.00 | 0.66 | 0.26 |

**Table 2**  Descriptive statistics of the physics network and particular realizations of its backbones, while other details are the same as in Table 1.

|  | Network | Convex Skeleton | Spanning Tree | Betweenness Edges | Embeddedness Edges |
|---|---|---|---|---|---|
| # Nodes | 425 | 425 | 425 | 425 | 425 |
| # Edges | 2,223 | 924 | 424 | 924 | 924 |
| % LCC | 100.0 | 100.0 | 100.0 | 97.6 | 20.9 |
| Degree | 10.46 | 4.35 | 2.00 | 4.35 | 4.35 |
| Distance | 3.62 | 8.75 | 9.53 | 3.81 | 2.31 |
| Assortativity | 0.07 | 0.41 | -0.26 | -0.20 | -0.05 |
| Clustering | 0.53 | 0.16 | 0.00 | 0.18 | 0.25 |
| Convexity | 0.32 | 0.91 | 1.00 | 0.48 | 0.09 |

**Table 3**  Descriptive statistics of the sociology network and particular realizations of its backbones, while other details are the same as in Table 1.

|  | Network | Convex Skeleton | Spanning Tree | Betweenness Edges | Embeddedness Edges |
|---|---|---|---|---|---|
| # Nodes | 145 | 145 | 145 | 145 | 145 |
| # Edges | 596 | 310 | 144 | 310 | 310 |
| % LCC | 100.0 | 100.0 | 100.0 | 97.9 | 35.9 |
| Degree | 8.22 | 4.28 | 1.99 | 4.28 | 4.28 |
| Distance | 3.23 | 6.02 | 5.65 | 3.41 | 2.65 |
| Assortativity | 0.23 | 0.62 | -0.23 | -0.23 | 0.09 |
| Clustering | 0.49 | 0.17 | 0.00 | 0.16 | 0.29 |
| Convexity | 0.48 | 0.94 | 1.00 | 0.49 | 0.22 |

**Table A.1**   Descriptive statistics of the mathematics network and particular realizations of its backbones, while other details are the same as in Table 1.

|  | Network | Convex Skeleton | Spanning Tree | Betweenness Edges | Embeddedness Edges |
|---|---|---|---|---|---|
| # Nodes | 167 | 167 | 167 | 167 | 167 |
| # Edges | 349 | 216 | 166 | 216 | 216 |
| % LCC | 100.0 | 100.0 | 100.0 | 98.8 | 30.5 |
| Degree | 4.18 | 2.59 | 1.99 | 2.59 | 2.59 |
| Distance | 4.65 | 9.04 | 9.11 | 4.75 | 2.93 |
| Assortativity | -0.06 | 0.14 | -0.30 | -0.20 | 0.07 |
| Clustering | 0.44 | 0.11 | 0.00 | 0.04 | 0.34 |
| Convexity | 0.64 | 0.99 | 1.00 | 0.75 | 0.21 |

**Table A.2**   Descriptive statistics of the economics network and particular realizations of its backbones, while other details are the same as in Table 1.

|  | Network | Convex Skeleton | Spanning Tree | Betweenness Edges | Embeddedness Edges |
|---|---|---|---|---|---|
| # Nodes | 414 | 414 | 414 | 414 | 414 |
| # Edges | 1,386 | 657 | 413 | 657 | 657 |
| % LCC | 100.0 | 100.0 | 100.0 | 92.3 | 44.2 |
| Degree | 6.70 | 3.17 | 2.00 | 3.17 | 3.17 |
| Distance | 4.25 | 11.07 | 8.84 | 4.57 | 4.60 |
| Assortativity | 0.05 | 0.49 | -0.25 | -0.08 | 0.06 |
| Clustering | 0.43 | 0.12 | 0.00 | 0.05 | 0.29 |
| Convexity | 0.38 | 0.98 | 1.00 | 0.48 | 0.29 |